\documentclass[fleqn,usenatbib]{mnras}

\usepackage{newtxtext}
\usepackage{xcolor}

\usepackage[T1]{fontenc}

\DeclareRobustCommand{\VAN}[3]{#2}
\let\VANthebibliography\thebibliography
\def\thebibliography{\DeclareRobustCommand{\VAN}[3]{##3}\VANthebibliography}

\newcommand{\Nparent}[1]{63,721}
\newcommand{\Nvar}[1]{706}
\newcommand{\Ndwarf}[1]{26}
\newcommand{\Nfinal}[1]{15}

\newcommand{\NWISE}[1]{706}
\newcommand{\NWISEAGN}[1]{389}
\newcommand{\NXray}[1]{105}
\newcommand{\Nsdss}[1]{?}

\newcommand{\RMSEmassvar}[1]{0.77}
\newcommand{\RMSEmassnonvar}[1]{1.25}
\newcommand{\RMSEzvar}[1]{0.35}
\newcommand{\RMSEznonvar}[1]{0.25}

\newcommand{\MgII}{Mg\,{\small II}}

\newcommand{\hbeta}{H{$\beta$}}

\newcommand{\OIII}{[O\,{\small III}]}

\usepackage{graphicx}	
\usepackage{amsmath}	
\usepackage{amssymb}	
\usepackage{caption}

\usepackage{eso-pic}

\AddToShipoutPictureBG*{%
  \AtPageUpperLeft{%
    \hspace{0.735\paperwidth}%
    \raisebox{-1.5\baselineskip}{%
      \makebox[0pt][l]{\textnormal{DES-2020-0555}}
}}}%

\AddToShipoutPictureBG*{%
  \AtPageUpperLeft{%
    \hspace{0.735\paperwidth}%
    \raisebox{-2.5\baselineskip}{%
      \makebox[0pt][l]{\textnormal{FERMILAB-PUB-20-103-AE}}
}}}%

\title[DES Deep Field Dwarf AGNs]{Dwarf AGNs from Optical Variability for the Origins of Seeds (DAVOS): Insights from the Dark Energy Survey Deep Fields}

\author[C. J. Burke et al.]{
Colin~J.~Burke,$^{1,2}$\thanks{E-mail: colinjb2@illinois.edu; xinliuxl@illinois.edu}
Xin~Liu,$^{1,3}$
Yue~Shen,$^{1,3}$
Kedar~A.~Phadke,$^{1}$
Qian~Yang,$^{1}$
Will~G.~Hartley,$^{4}$
\newauthor
Ian~Harrison,$^{5}$
Antonella~Palmese,$^{6,7}$
Hengxiao~Guo,$^{1,3,8}$
Kaiwen~Zhang,$^{9}$
Richard~Kron,$^{6,10}$
\newauthor
David~J.~Turner,$^{11}$
Paul~A.~Giles,$^{11}$
Christopher~Lidman,$^{12}$
Yu-Ching~Chen,$^{1}$
Robert~A.~Gruendl,$^{1,2}$
\newauthor
Ami~Choi,$^{13}$
Alexandra~Amon,$^{14}$
Erin~Sheldon,$^{15}$
M.~Aguena,$^{16}$
S.~Allam,$^{6}$
F.~Andrade-Oliveira,$^{17,16}$
\newauthor
D.~Bacon,$^{18}$
E.~Bertin,$^{19,20}$
D.~Brooks,$^{21}$
A.~Carnero~Rosell,$^{16}$
M.~Carrasco~Kind,$^{1,2}$
J.~Carretero,$^{22}$
\newauthor
C.~Conselice,$^{5,23}$
M.~Costanzi,$^{24,25,26}$
L.~N.~da Costa,$^{16,27}$
M.~E.~S.~Pereira,$^{28,29}$
T.~M.~Davis,$^{30}$
\newauthor
J.~De~Vicente,$^{31}$
S.~Desai,$^{32}$
H.~T.~Diehl,$^{6}$
S.~Everett,$^{33}$
I.~Ferrero,$^{34}$
B.~Flaugher,$^{6}$
J.~Garc\'ia-Bellido,$^{35}$
\newauthor
E.~Gaztanaga,$^{36,37}$
D.~Gruen,$^{38}$
J.~Gschwend,$^{16,27}$
G.~Gutierrez,$^{6}$
S.~R.~Hinton,$^{30}$
D.~L.~Hollowood,$^{33}$
\newauthor
K.~Honscheid,$^{13,40}$
B.~Hoyle,$^{38}$
D.~J.~James,$^{41}$
K.~Kuehn,$^{42,43}$
M.~A.~G.~Maia,$^{16,27}$
J.~L.~Marshall,$^{44}$
\newauthor
F.~Menanteau,$^{1,2}$
R.~Miquel,$^{45,22}$
R.~Morgan,$^{46}$
F.~Paz-Chinch\'{o}n,$^{2,47}$
A.~Pieres,$^{16,27}$
A.~A.~Plazas~Malag\'on,$^{48}$
\newauthor
K.~Reil,$^{49}$
A.~K.~Romer,$^{11}$
E.~Sanchez,$^{31}$
M.~Schubnell,$^{28}$
S.~Serrano,$^{36,37}$
I.~Sevilla-Noarbe,$^{31}$
M.~Smith,$^{50}$
\newauthor
E.~Suchyta,$^{51}$
G.~Tarle,$^{28}$
D.~Thomas,$^{18}$
C.~To,$^{52,53,49}$
T.~N.~Varga,$^{54,55}$
and R.D.~Wilkinson$^{11}$
\newauthor
(DES Collaboration)
\\
\\
\emph{\normalsize The author's affiliations are shown at the end of this paper.}
}

\date{Accepted XXX. Received YYY; in original form ZZZ}

\pubyear{2021}

\begin{document}
\label{firstpage}
\pagerange{\pageref{firstpage}--\pageref{lastpage}}
\maketitle

\begin{abstract}
We present a sample of \Nvar{}, $z < 1.5$ active galactic nuclei (AGNs) selected from optical photometric variability in three of the Dark Energy Survey (DES) deep fields (E2, C3, and X3) over an area of 4.64 deg$^2$. We construct light curves using difference imaging aperture photometry for resolved sources and non-difference imaging PSF photometry for unresolved sources, respectively, and characterize the variability significance. Our DES light curves have a mean cadence of 7 days, a 6 year baseline, and a single-epoch imaging depth of up to $g \sim 24.5$. Using spectral energy distribution (SED) fitting, we find \Ndwarf{} out of total \Nvar{} variable galaxies are consistent with dwarf galaxies with a reliable stellar mass estimate ($M_{\ast}<10^{9.5}\ M_\odot$; median photometric redshift of 0.9). We were able to constrain rapid characteristic variability timescales ($\sim$ weeks) using the DES light curves in \Nfinal{} dwarf AGN candidates (a subset of our variable AGN candidates) at a median photometric redshift of 0.4. This rapid variability is consistent with their low black hole masses. We confirm the low-mass AGN nature of one source with a high S/N optical spectrum. We publish our catalog, optical light curves, and supplementary data, such as X-ray properties and optical spectra, when available. We measure a variable AGN fraction versus stellar mass and compare to results from a forward model. This work demonstrates the feasibility of optical variability to identify AGNs with lower black hole masses in deep fields, which may be more ``pristine'' analogs of supermassive black hole seeds.

\end{abstract}




\section{Introduction} \label{sec:intro}

Virtually every massive galaxy contains a supermassive black hole (SMBH) in its center \citep{Kormendy1995}. There is growing evidence for the existence of intermediate-mass BHs (IMBHs, $M_{\bullet} = 10^2 \sim 10^6M_{\odot}$; \citealt{Greene2020}) in dwarf galaxies beyond the handful of well-studied examples: NGC 4395 \citep{Filippenko2003}, Pox 52 \citep{Barth2004}, Henize 2-10 \citep{Reines2015} and RGG 118 \citep{Baldassare2015}. The recent discovery of the gravitational wave transient GW190521 with a merger remnant mass of $142^{+28}_{-16}$ $M_{\odot}$ \citep{LIGOVirgo2020} provides the strongest evidence for IMBHs. However, the occupation fraction of black holes (BHs) in the dwarf galaxy regime remains poorly constrained \citep{Greene2020}. 

SMBHs as massive as several billion solar masses were already formed when the universe was only a few hundred Myr old \citep[e.g.][]{Fan2001,Wu2015,Banados2018,Wang2021}. How they were able to form so quickly is an outstanding question in cosmology \citep{Volonteri2010,Inayoshi2020}. At least three channels have been proposed for the formation of the seeds of SMBHs: Pop. III stellar remnants \citep[e.g.][]{Madau2001}, direct collapse \citep[e.g.][]{Haehnelt1993,Bromm2003,Begelman2006}, or star cluster evolution \citep[e.g.][]{Gurkan2004,PortegiesZwart2004}. The occupation fraction of BHs in local dwarf galaxies (i.e. $M_{\ast} < 10^{10} M_{\odot}$; \citealt{Greene2020}) and their mass functions traces the SMBH seeding mechanism at high redshifts \citep[e.g.][]{Greene2012,Reines2016}. The occupation function of BHs in ultra-dwarf ($M_{\ast} = 10^5 \sim 10^6 M_{\odot}$) galaxies is important for understanding the origin of some LIGO binary BHs \citep{Palmese2020a}. However, systematic approaches to finding such dwarf active galactic nuclei (AGNs) have only recently begun.

For example, deep X-ray surveys can be used to identify low-mass and low-luminosity AGNs at low and intermediate redshifts \citep{Fiore2012,Young2012,Civano2012,Luo2017,Xue2017}. However, these surveys are expensive and often plagued by contamination from X-ray binaries. Radio searches have also identified low-mass AGNs in star-forming dwarf galaxies \citep{Mezcua2019,Reines2020}, although they are subject to the
low detection rate of radio cores of AGNs. Alternatively, optical color selection is much less expensive but is biased against smaller BHs and/or lower Eddington ratios. Optical emission line selection, such as with BPT diagram diagnostics \citep{Baldwin1981,Veilleux1987}, is known to miss AGNs with line ratios dominated by star formation \citep{Baldassare2016,Agostino2019}, particularly in low-metallicity \citep{Groves2006} and low-mass galaxies without sufficient spectral resolution \citep{Trump2015}, and because of the dilution from star-forming regions within the spectral aperture in low-mass galaxies \citep{Mezcua2020,Yan2012}. Furthermore, the standard optical narrow emission line diagnostics used to identify AGNs may fail when the BH mass falls below ${\sim}10^4\ \rm M_\odot$ for highly accreting IMBHs and for radiatively inefficient IMBHs with active star formation, because the enhanced high-energy emission from IMBHs could result in a more extended partially ionized zone compared with models for SMBHs, producing a net decrease in the predicted [O~\textsc{iii}]$/$H$\beta$ and [N~\textsc{ii}]$/$H$\alpha$ emission line ratios \citep{Cann2019}. Recently, dwarf AGNs have been identified using coronal line emission signatures \citep{Cann2021,Molina2021}, but this requires high quality infrared spectra.

Compared to other techniques, variability searches should be more sensitive to AGNs with lower Eddington ratios given the anti-correlation between Eddington ratio and optical variability \citep{MacLeod2010,Rumbaugh2018}. The optical variability-selection technique for unobscured AGNs and quasars is well-established \citep{Trevese2008,Butler2011,Kumar2015,Cartier2015,DeCicco2015,Tie2017,SanchezSaez2018,DeCicco2019,Pouliasis2019,Poulain2020,Costa2020,Kimura2020}. Also see \citet{Elmer2020} and \citet{Secrest2020} for recent studies based on near-infrared (NIR) and mid-infrared (MIR) variabilities. Variability results in an incomplete selection, missing optically-obscured AGNs or those with bright host galaxies that dilute the variability from the accretion disk. The selection rates are expected to depend on the sensitivity/photometric precision of the survey and the exact selection criteria used \citep{Burke2022}. However, using variability as a complementary AGN selection technique to identify dwarf AGNs is a relatively new technique \citep{Baldassare2018,Martinez-Palomera2020,Baldassare2020,Guo2020,Ward2021b}.

In this work, we perform a systematic search for variable AGN using Dark Energy Survey \citep[DES;][]{DarkEnergySurveyCollaboration2016} deep field imaging \citep{Hartley2020}. We choose the DES deep fields because of the exceptional depth ($g\sim24.6$), $\sim$7 day cadence with a total baseline of $\sim$6 years, and availability of multi-wavelength imaging and spectroscopy. Using these data, we are able to identify optically variable AGN candidates in dwarf galaxies to $z\sim1.5$ for the first time.

This paper is organized as follows. In \S\ref{sec:data} we describe the DES observations, our methods for constructing light curves, and our variability-selection procedure, in \S\ref{sec:results} we present our catalog of variability-selected dwarf AGNs and study our AGN detection fraction, in \S\ref{sec:discussion} we compare our results to previous works, in \S\ref{sec:conclusions} we summarize our new findings and conclude.

\section{Observations and Data Analysis}\label{sec:data}

\subsection{The Dark Energy Survey}\label{sec:des}

The DES (Jan 2013–Jan 2019) was a wide-area $\sim5000$ deg$^2$ survey of the southern galactic cap in the \emph{grizY} bands. It used the Dark Energy Camera \citep{Flaugher2015,Bernstein2017} with a 2.2 degree diameter field of view mounted at the prime focus of the Victor M. Blanco 4-m telescope on Cerro Tololo in Chile. The data quality varies due to seeing and weather variations. The DES absolute photometric calibration has been tied to the spectrophotometric Hubble CALSPEC standard star C26202 and has been placed on the AB system \citep{Oke1983}. In addition to the wide-area survey, the DES contains a 27 deg$^2$ multi-epoch survey to search for Type Ia supernovae (SNe) called DES-SN \citep{Kessler2015}. DES-SN is composed of 10 DES fields each with a uniform cadence of about 7 days in the \emph{griz} bands during the observing season. DES-SN operated during the ``science verification'' -- year 5 (SV--Y5) seasons (6 year total baseline). 

\subsection{The Dark Energy Survey Deep Fields}\label{sec:deepfields}

\begin{table*}
\centering
\caption{Summary of DES deep fields used in this work. The SN-E3 field is referred to as a ``shallow'' field by \citet{Kessler2015} but the coadd photometry from \citet{Hartley2020} is built to the same depth as the ``deep'' fields, SN-C3 and SN-X3. Column \#4 refers to the single-epoch limiting PSF magnitude depth where the detection efficiency has fallen to 50\% \citep{Kessler2015}. The median number of epochs refers to our light curves after requiring $N_{\rm{epoch}}>100$.
\label{tab:deepfields}}
\begin{tabular}{ccccc}
\hline
Field & RA [deg (hh:mm:ss)] & dec [deg (hh:mm:ss)] & Limiting $g$-mag & Median $N_{\rm epochs}$ \\
\hline
 SN-C3     & 52.6484 (03:30:35.6)  & $-$28.1000 ($-$28:06:00.0) & 24.5 & 364 \\ 
 SN-E2     & 9.5000 (00:38:00.0)   & $-$43.9980 ($-$43:59:52.8) & 23.5 & 120 \\
 SN-X3     & 36.4500 (02:25:48.0)  & $-$4.6000 ($-$04:36:00.0)  & 24.5 & 163 \\
\hline
\end{tabular}
\end{table*}

In this work, we restrict our analysis to three of the 11 (composed of 10 DES-SN fields plus the COSMOS field) DES deep fields (SN-E2, SN-C3, SN-X3) with weekly cadence from the DES-SN program \citep{Kessler2015} and with 8-band ($ugrizJHK_{S}$) deblended, stacked model-based photometry from \citet{Hartley2020} (Table~\ref{tab:deepfields}). These fields overlap with the European Large Area ISO Survey \citep{Oliver2000}, the \emph{Chandra} Deep Field-South \citep{Luo2017}, and \emph{XMM} Large Scale Structure survey fields \citep{Garcet2007}, respectively. Supplementary DECam \emph{u}-band imaging was obtained in these fields. Additional $JHK_{S}$ imaging data are from the VIDEO (overlaps with SN-E2, SN-C3, and SN-X3; \citealt{Jarvis2013}) and UltraVISTA (overlaps with COSMOS; \citealt{McCracken2012}) surveys, and the final deblended catalog is built to a uniform depth of $i=25$. The total area of the fields with NIR overlap is 4.64 deg$^2$ after masking bright stars and artifacts. We will leverage the deep 8-band color information for star-galaxy separation and stellar mass estimates.

\subsection{Star-Galaxy Separation} \label{sec:deepfieldcatalog}

\begin{figure}
\includegraphics[width=0.5\textwidth]{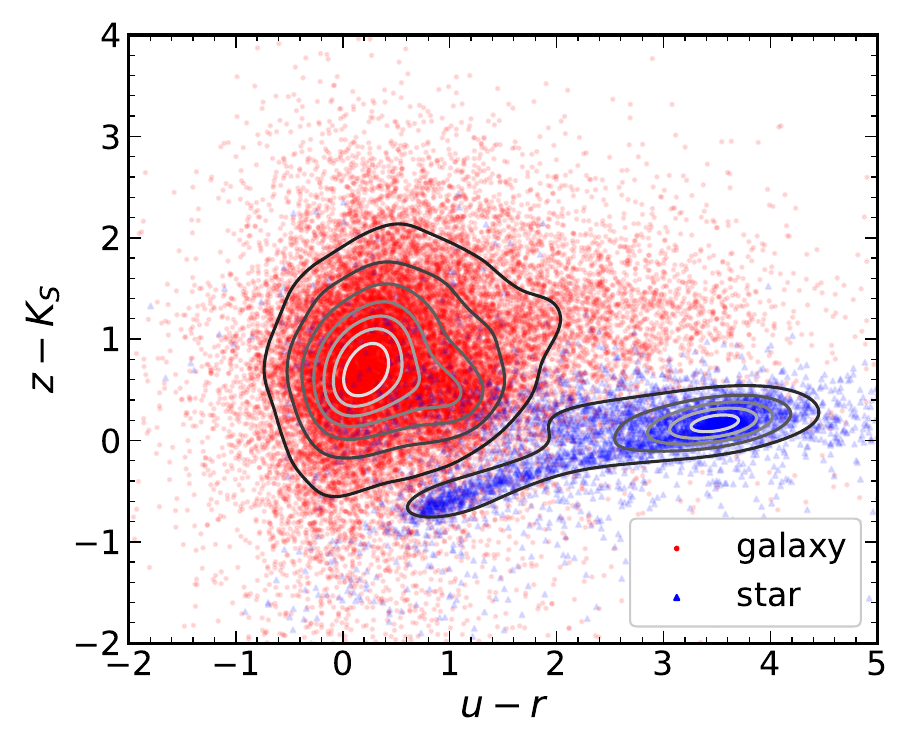}

\caption{Color-color plot demonstrating the star-galaxy classifier using the machine learning method described in \S\ref{sec:deepfieldcatalog}. For clarity, a subset of all objects in SN-C3 are plotted under density contours. \label{fig:stargalaxy}}
\end{figure}

\begin{figure}
\includegraphics[width=0.5\textwidth]{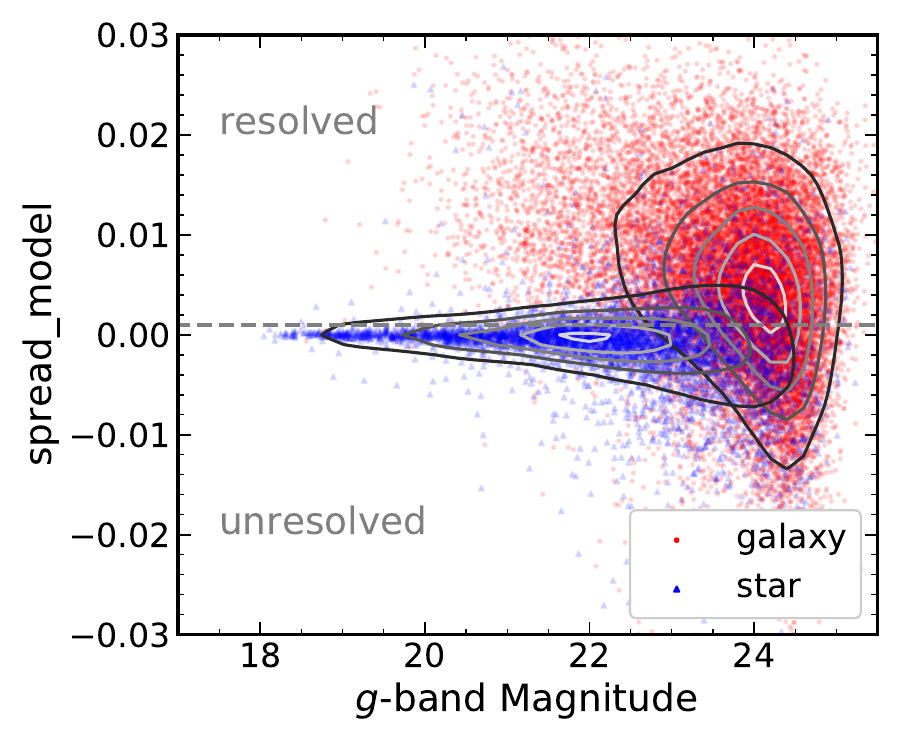}

\caption{Spread model versus median $g$-band PSF magnitude for galaxies (red circles) and stars (blue triangles) classified using the machine learning classifier described in \S\ref{sec:deepfieldcatalog}. The dashed gray line shows the cut at \texttt{spread\_model} $=0.001$ between resolved and unresolved sources. For clarity, a subset of the total data points in SN-C3 are plotted under density contours. \label{fig:spreadmodel}}
\end{figure}

Star-galaxy separation is performed using a supervised machine learning classifier trained on DES-COSMOS \emph{ugriz} and UltraVISTA survey $JHK_{S}$ band imaging using the \emph{Hubble Space Telescope} morphological star-galaxy classifications of \citet{Leauthaud2007} as the ground truth. The trained classifier is then applied to the DES-SN fields, with additional validation shown in \citet{Hartley2020}. A $k$-nearest neighbors method is used, which yields a purity and completeness of $\sim 99\%$ or better \citep{Hartley2020}. A color-color plot demonstrating the star-galaxy classifier is shown in Figure~\ref{fig:stargalaxy}.

\subsection{Light Curve Construction} \label{sec:pipeline}

\begin{figure}
\includegraphics[width=0.5\textwidth]{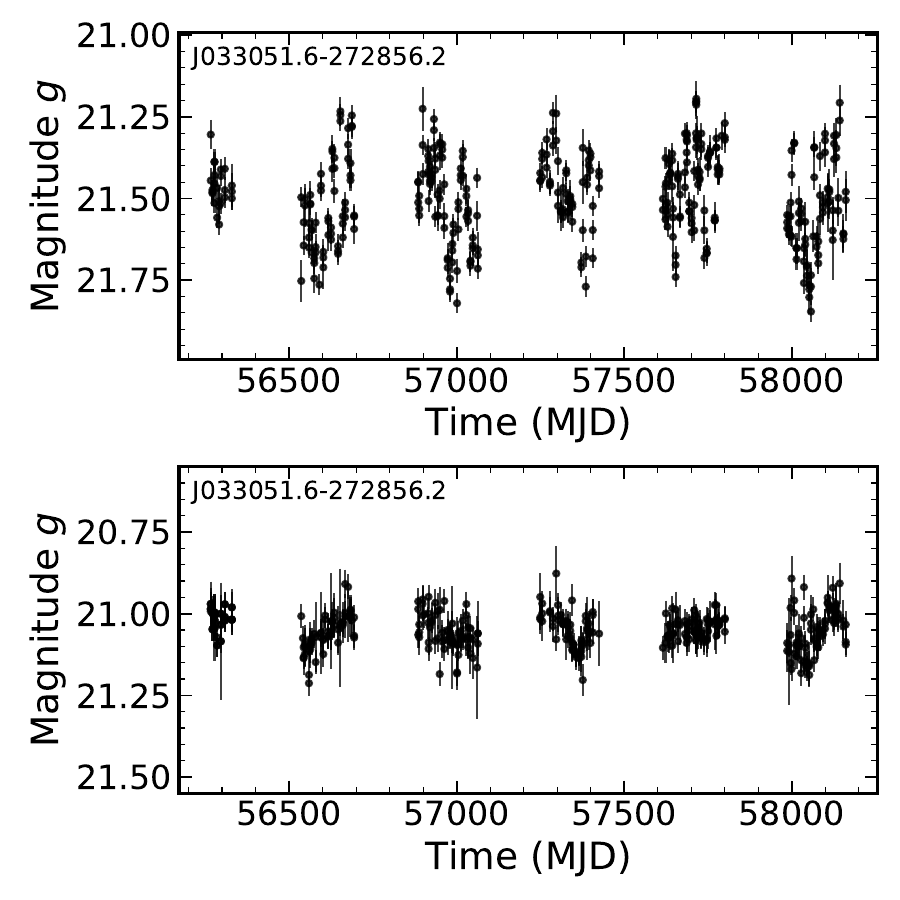}

\caption{Example PSF (\emph{top}) and DIA (\emph{bottom}) light curves of a resolved variable galaxy. The PSF light curve includes additional false variability due to seeing variations. Significant intrinsic variability is still detected in this source using DIA photometry. The slight difference in magnitude for resolved sources is due to the larger area enclosed in the DIA aperture. \label{fig:diaex}}
\end{figure}

\begin{figure}
\includegraphics[width=0.5\textwidth]{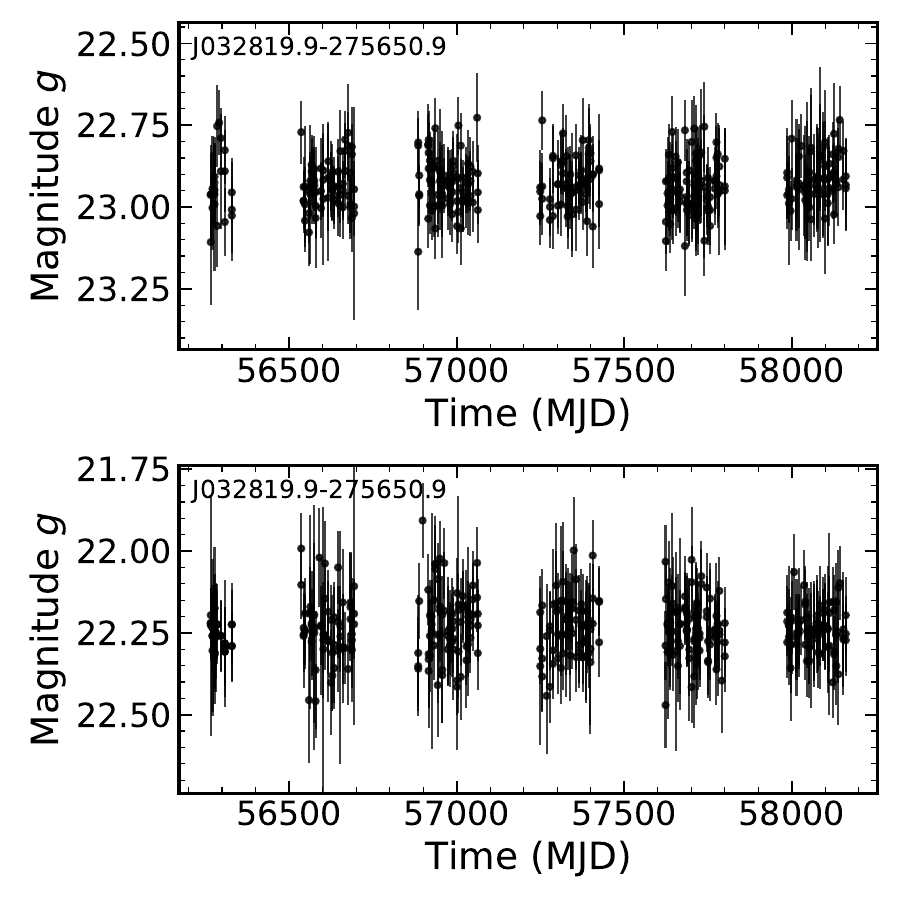}

\caption{Example PSF (\emph{top}) and DIA (\emph{bottom}) light curves of an unresolved non-variable star. The slight difference in magnitude is due to the larger area enclosed in the DIA aperture. \label{fig:diaexstar}}
\end{figure}

We construct light curves using \emph{g}-band point spread function (PSF) magnitudes for unresolved sources. For resolved sources, we use aperture-based difference imaging analysis (DIA) magnitudes. While light curves in other bands can be useful, it is computationally expensive to re-compute difference images in all bands. Given AGNs generally are more variable in bluer bands and DES-SN did not perform \emph{u}-band imaging, we chose to restrict our variability selection to the \emph{g}-band. In addition, the accretion disk SED is expected to shift into the bluer/UV part of the spectrum at lower black hole masses (e.g., \citealt{Cann2019}). Futhermore, differences between variability timescales between bands is small, scaling like $\lambda^{0.17}$ \citep{MacLeod2010,Suberlak2021}. To determine whether to use PSF or DIA photometry, we use the \texttt{spread\_model} estimator to separate resolved and unresolved sources (e.g., \citealt{Desai2012,Soumagnac2015}). This estimator is the normalized simplified linear discriminant between a local PSF model $\tilde{\phi}$ and an extended model $\tilde{G}$:
\begin{equation}
    \texttt{spread\_model} = \frac{\tilde{G}^T W p}{\tilde{\phi}^T W p} - \frac{\tilde{G}^T W \tilde{\phi}}{\tilde{\phi}^T W \tilde{\phi}},
\end{equation}
where $\tilde{G}$ is the local PSF model convolved with a circular exponential disk model with scale length $1/16$th of the full width at half-maximum of the PSF model, $p$ is the image vector centered on the source, and $W$ is a weight matrix constant along the diagonal. We chose the threshold \texttt{spread\_model} $=0.001$ (i.e., use DIA magnitudes if \texttt{spread\_model} $>0.001$, otherwise use PSF magnitudes). We use the median \texttt{spread\_model} of all measurements for each source to select a pure sample of resolved sources. The threshold is shown in Figure~\ref{fig:spreadmodel}. As a demonstration of our DIA photometry, we also show example PSF and DIA light curves of a resolved galaxy in Figure~\ref{fig:diaex} and for an unresolved non-variable star in Figure~\ref{fig:diaexstar}. DIA allows us to measure variability from the central AGN (a point source) after subtracting the flux from the non-variable extended host galaxy. Simple aperture photometry is generally inadequate for resolved sources, because seeing variations can contaminate the aperture with varying fractions of light from the host galaxy.

Our difference imaging pipeline is similar to the DES-SN pipeline of \cite{Kessler2015}, used to identify Type Ia SNe and other transients in the DES-SN fields. However, we need to re-compute the difference images across all observing seasons with a single template image to achieve a consistent zero-point light curve. In contrast, the DES-SN difference images were computed using a different template in each season. The DIA pipeline uses the \textsc{hotpants} code \citep{Becker2013}, which follows the algorithm described in \citet{Alard1998} and \citet{Alard2000}. We build a PSF kernel across each single-epoch science image and convolve the template to match the PSF in the science image. \textsc{hotpants} works by minimizing the equation,
\begin{equation}
    -2 \log L = \sum_i \left( \left[T \otimes K \right](x_i,y_i) - I(x_i,y_i) \right)^2
\end{equation}
where $T$ is the template frame, $K$ is the PSF kernel, $I$ is the science frame, and $\otimes$ denotes convolution. We assume that the kernel $K$ can be decomposed into Gaussian basis functions which are allowed to vary on differing spatial orders. This takes the form,
\begin{equation}
    K(u,v)=\sum_n a_nK_n(u,v),
\end{equation}
where $K_n(u,v)=e^{-\left(u^2+v^2\right)/2\sigma_k^2}u^iv^j$ and $n=(i,j,k)$. The spatial order is confined to the size of kernel. 

We adapted the pipeline from \cite{Kessler2015} to produce a single template used for all DES-SN seasons (to achieve a constant zero point), and perform forced photometry centered on detections in the template image. We follow the criteria of \cite{Kessler2015} for creating the template image, selecting Y3 images with sky noise $\sigma_{\text{sky}} < 2.5\ \sigma_{\text{sky, min}}$. After this, we use up to 10 images with the smallest PSF. The template and image subtraction are done on a per-CCD basis ($2048\times4096$ pixels). We use \textsc{SWarp} to create the template coadds and reproject each science image to the template WCS. We use \textsc{SExtractor} \citep{Bertin1996} in double image mode to perform forced photometry on the template and difference image. We use a 5$\sigma$ threshold for detection in the template image. We use a circular aperture of 5$^{\prime\prime}$ in diameter to be larger than the seeing disk. This restriction to nuclear variability precludes us from detecting off-nuclear (recoiling or wandering) SMBHs \citep{Blecha2016,Reines2020,Ward2021} $\gtrsim 2.5^{\prime\prime}$ from a galaxy's centroid. Although it is possible to measure the position of the variability using the difference frames, image artifacts caused by small astrometric misalignments makes this difficult in practice. Therefore, we leave detecting off-nuclear AGNs to future studies. 

\subsection{Variability-Selection Procedure} \label{sec:selection}

\subsubsection{Variability Significance}

After constructing light curves and determining whether to use PSF or DIA magnitudes, we perform outlier rejection on each light curve using a sliding window approach. We use a window size of 150 days and a $3\sigma$ rejection threshold where $\sigma=1.4826\ \rm{MAD}$, where $\rm{MAD}$ is the median absolute deviation. We also empirically correct the photometric uncertainties for systematics following the method of \citet{Sesar2007}, as detailed in Appendix~\ref{sec:errorbarcorrection}. We use the $\chi^2$-based maximum-likelihood estimator from \citet{Shen2019} to estimate the intrinsic variability of sources, as described below.

For a light curve with photometry $X_i$ and measurement error $\sigma_i$ and unknown excess variance $\sigma_0^2$ from intrinsic variability, we have:
\begin{equation}
    \text{Var}[X_i] = \sigma_0^2+\sigma_i^2 = \frac{\sigma_0^2}{g_i},
\end{equation}
where
\begin{equation}
    g_i \equiv \frac{\sigma_0^2}{\sigma_0^2+\sigma_i^2} = \frac{1}{1+ (\sigma_i/\sigma_0)^2}
\end{equation}
quantifies the ``goodness'' of $X_i$ for measuring $\sigma_0^2$. $g_i$ varies from 0 for points with $\sigma_i \gg \sigma_0$ to 1 for points with $\sigma_i \ll \sigma_0$. The sum of $g_i$ over all data points then provides a goodness of measuring the intrinsic variability using the time series and approaches the total number of data points in the limit of $\sigma_i \ll \sigma_0$.

The likelihood function given $X_i$ and a constant flux model of $\mu=\left< X_i \right>$ with both measurement errors and intrinsic variance is:
\begin{equation}
    -2 \log L = \sum_{i=1}^N \frac{(X_i-\mu)^2}{\sigma_0^2+\sigma_i^2} + \sum_{i=1}^N \log (\sigma_0^2+\sigma_i^2).
\end{equation}
Minimizing the likelihood function, we obtain an estimate of $\sigma_0$ as
\begin{equation}
\begin{split}
    \hat\sigma_0^2 = \frac{\sum (X_i-\mu)^2 g_i^2}{\sum g_i}, \\
    \text{Var}[\sigma_0^2] = \frac{\hat\sigma_0^4}{\sum g_i \frac{\sum (X_i-\mu)^2 g_i^3}{\sum (X_i-\mu)^2 g_i^2} - \sum g_i^2/2 }.
    \label{eq:rms_mle}
\end{split}
\end{equation}
To estimate the value of $\mu$, we use the optimal weights of the photometry based on $\sigma_i$ and $\sigma_0$:
\begin{equation}
    \hat\mu = \frac{\sum \frac{X_i}{\sigma_0^2+\sigma_i^2}}{\sum \frac{1}{\sigma_0^2+\sigma_i^2}} = \frac{\sum X_i g_i}{\sum g_i},\; \rm{Var}[\mu] = \frac{\sigma_0^2}{\sum g_i}.
    \label{eq:mu_mle}
\end{equation}
Equations \ref{eq:rms_mle} and \ref{eq:mu_mle} are solved iteratively. We have neglected the (usually small) covariance between $\hat\mu$ and $\hat\sigma_0^2$. We define signal-to-noise ratio estimator as,
\begin{equation}
    \text{SNR} = \frac{\hat\sigma_0}{ \text{RMS}[\sigma_0]}.
    \label{eq:snr}
\end{equation}
As noted by \citet{Shen2019}, SNR saturates near $\sqrt{2(N_{\rm{epoch}}-1)}$ where $N_{\rm{epoch}}$ is the number of epochs in the light curve. To classify a source as variable, we require $N_{\rm{epoch}}>100$ and $\text{SNR}>3$. We will compare this estimator with other variability statistics later to justify the adopted threshold.

\begin{figure*}
\includegraphics[width=0.85\textwidth]{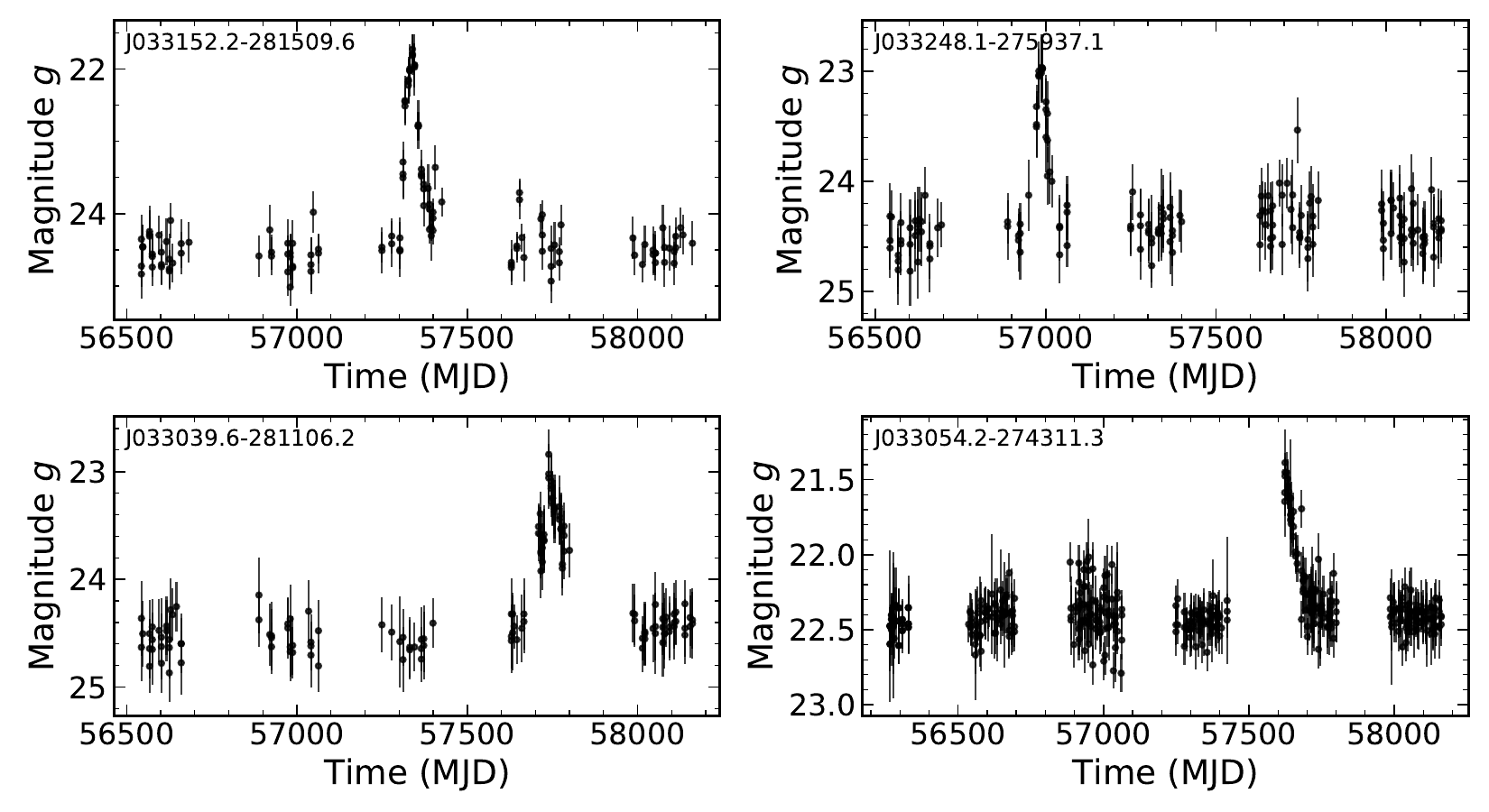}
\centering
\caption{Examples of clear non-AGN-like transients rejected by our method described in \S\ref{sec:transient}, including DES J033152.2-281509.6 (known superluminous SN DES15C3hav; \citealt{Angus2019}). \label{fig:transients}}
\end{figure*}

\subsubsection{Transient Rejection}
\label{sec:transient}

To reject flaring transients from our sample (e.g., SNe, tidal disruption events, microlensing events) with timescales of less than $\sim 1$ year, we determine if the variability is confined to only one light curve season. For each of the six seasons $N_i$ in the light curve, we compute the SNR using the light curve data in the other five seasons without the data from season $N_i$. If any of these seasonal SNR values falls below 2, we flag the light curve as a possible transient and exclude it from our analysis. Examples of flagged light curves are shown in Figure~\ref{fig:transients}. Still, rare long-duration optical stellar transients, such as outbursts of massive stars, can mimic AGN variability \citep{Burke2020}. Therefore, we must be cautious before confirming the AGN nature of our candidates.

\subsubsection{AGN-like Variability}

To further increase the purity of our sample, we use the auto-correlation information of the light curves to reject light curves with spurious variability which appear as white noise. This contrasts to AGN light curves which show a correlated behavior, commonly modeled as a damped random walk \citep{MacLeod2010,Kelly2011}. Specifically, we use the Ljung–Box test under the null hypothesis that the light curve data are independently distributed with time \citep{LjungBox1978}. The Ljung–Box test is a portmanteau test, which does not evaluate the light curve against a particular model of intrinsic variability. We convert the test statistic to a significance that the light curve is ``AGN-like'' without any particular priors on the structure function or model assumptions. We denote this quantity as $\sigma_{\rm{LB}}$. Finally, we note that $\sigma_{\rm{LB}}$ will be small for AGNs that vary predominately on timescales less than the $\sim 7$ day DES-SN cadence. We discuss this in more detail and the possible selection biases this may induce below.

\subsubsection{Comparison to a Quasar-Selection Method}
\label{sec:qsofit}

\begin{figure*}
\includegraphics[width=0.85\textwidth]{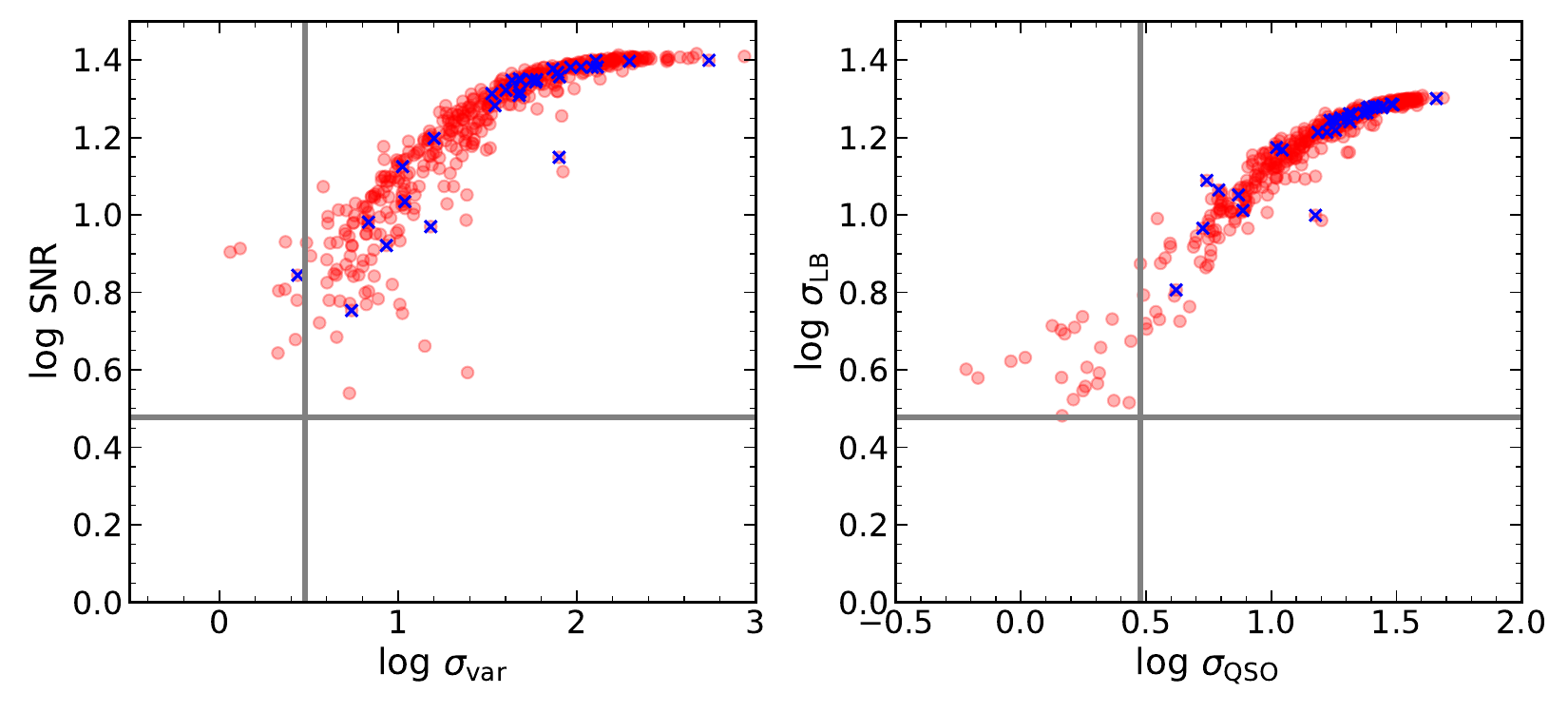}
\centering
\caption{Comparison of variability significance criteria used in this work to the method of \citet{Butler2011} on DES light curves of galaxies. We plot the \citet{Shen2019} $\chi^2$-based estimator ($\log \rm{SNR}$) used in this work versus the simple $\chi^2/\nu$ variability significance ($\sigma_{\rm{var}}$) of \citet{Butler2011} (\emph{left}). Galaxies with robustly-estimated variability are located in the upper-right hand corner of the figure panel. We also plot the Ljung–Box test significance ($\sigma_{\rm{LB}}$) used in this work versus the parametric quasar significance ($\sigma_{\rm{QSO}}$) of \citet{Butler2011} (\emph{right}). Galaxies with AGN-like, correlated variability are located in the upper-right corner of the figure panel. The solid gray lines indicate 3$\sigma$ thresholds; sources which pass both variability tests appear in the upper-right hand boxes in each panel. For clarity, only SN-C3 sources with ${\rm{SNR}} > 3$ are shown. Sources with \emph{Chandra} X-ray detections are shown as blue cross symbols. 
\label{fig:qsofit}}
\end{figure*}

As described above, our non-parametric SNR and $\sigma_{\rm{LB}}$ criteria do not evaluate the light curve against a particular model of intrinsic variability. This is necessary to avoid selection biases associated with particular model parameters. In contrast, the \citet{Butler2011} method is based on two critera: a $\chi^2$-test variability estimator ($\sigma_{\rm{var}}$) and a model significance ($\sigma_{\rm{QSO}}$) evaluated against a parameterization of the ensemble quasar structure function as a function of apparent magnitude using quasars in SDSS Stripe 82. It is unclear if this parameterization is optimal for selection of dwarf AGNs, because the structure function of dwarf AGNs is not well-studied. In addition, the fractional contamination of the host galaxy flux is generally larger for dwarf AGNs, which have lower AGN luminosities than quasars. This means the parameterization as a function of apparent magnitude may not be valid for the sources we are interested in. However, \citet{Baldassare2018,Baldassare2020} found 0.25\% to 1.0\% of $z<0.15$ dwarf galaxies had a detectable variable AGN, depending on the light curve baseline, using the \citet{Butler2011} method.

To better study the efficacy of both techniques, we compare our variability estimator to the \citet{Butler2011} selection method as implemented in the \textsc{qso\_fit} code\footnote{\url{http://butler.lab.asu.edu/qso_selection/index.html}}. The comparison is shown in Figure~\ref{fig:qsofit}. After removing transient sources, we find that our SNR metric is well-correlated with $\sigma_{\rm{var}}$ for variable sources (${\rm{SNR}} > 3$). In addition, we find that $\sigma_{\rm{LB}}$ and $\sigma_{\rm{QSO}}$ are well-correlated for variable sources (${\rm{SNR}} > 3$). This may imply that the AGNs in our sample have close-enough structure functions to normal quasars given the sensitivity of the \citet{Butler2011} $\sigma_{\rm{QSO}}$ test. Nevertheless, this comparison validates our variability-selection procedure, which has fewer assumptions about the intrinsic properties of AGN variability.

\subsection{Photometric Redshifts}
\label{sec:photoz}

We use the method of \citet{Yang2017}, which is trained on both non-AGN galaxies and AGNs, to determine the photometric redshifts $z_{\rm{ph}}$ of our sources using the available optical/NIR photometry from \citet{Hartley2020}. We start with sources classified as galaxies from the DES deep field $k$-nearest neighbors classifier (as opposed to stars). Then, we use the template fitting method described in \citet{Yang2017} to further classify sources as galaxies or quasars. Because we are looking for variable dwarf (low-luminosity) AGNs, we consider both quasar and galaxy classes. After determining the classification, we obtain $z_{\rm{ph}}$ values for each source using the asymmetries in the relative flux distributions as a function of redshift and magnitude \citep{Yang2017}, where source fluxes were measured from the coadded photometry. The procedure is identical for both variable and non-variable galaxies/quasars, because we are interested in measuring the variable fraction as a function of stellar mass. Throughout this work, we adopt these $z_{\rm{ph}}$ values for our sources for the stellar mass estimation and analysis, described below.

\subsection{Stellar Mass Estimation}
\label{sec:stellarmass}

\begin{figure}
\centering
{\includegraphics[width=0.48\textwidth]{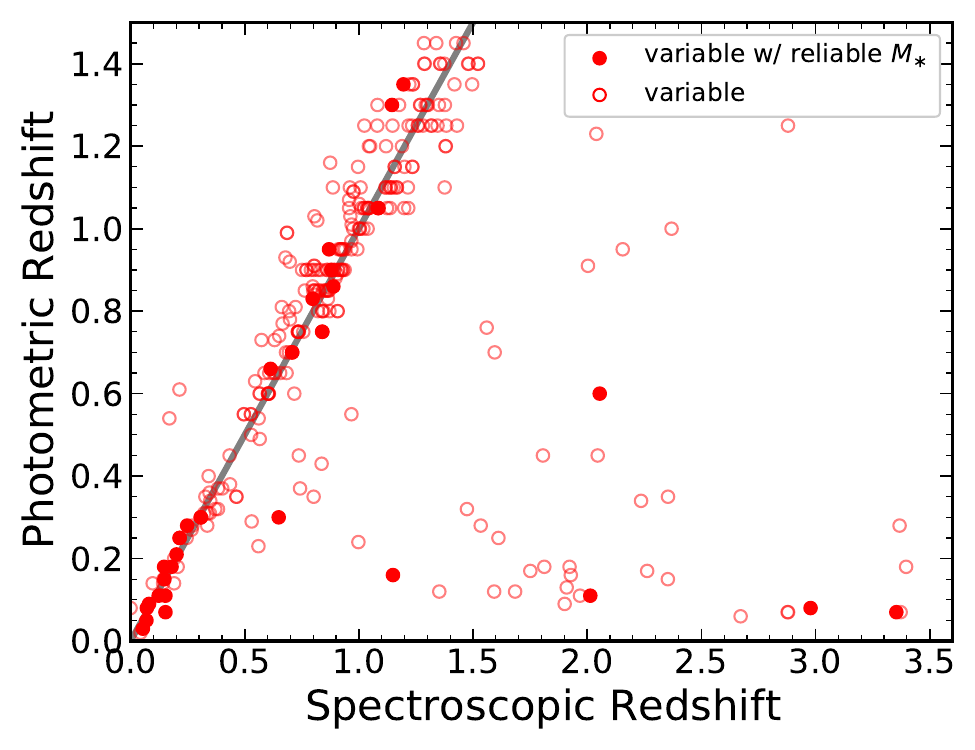}
}
\caption{We show the photometric redshift versus spectroscopic redshift (when available) for variable galaxies in our sample (red circles). The solid gray line is the $y=x$ line. Catastrophic photo-z failures in the lower right hand corner of the right panel are likely to be incorrectly identified as low stellar-mass galaxies by \textsc{cigale}. This is due to the degeneracy between the colors of low-$z$ star-forming galaxies and high-$z$ quasars. To remove most of these quasar interlopers, we reject sources that can be fit well with an AGN-dominated SED model, leaving sources with ``reliable'' stellar mass estimates (solid red circles; see~\S\ref{sec:stellarmass}). \label{fig:resolved_frac}}
\end{figure}

\begin{figure*}
\centering
{\includegraphics[width=0.8\textwidth]{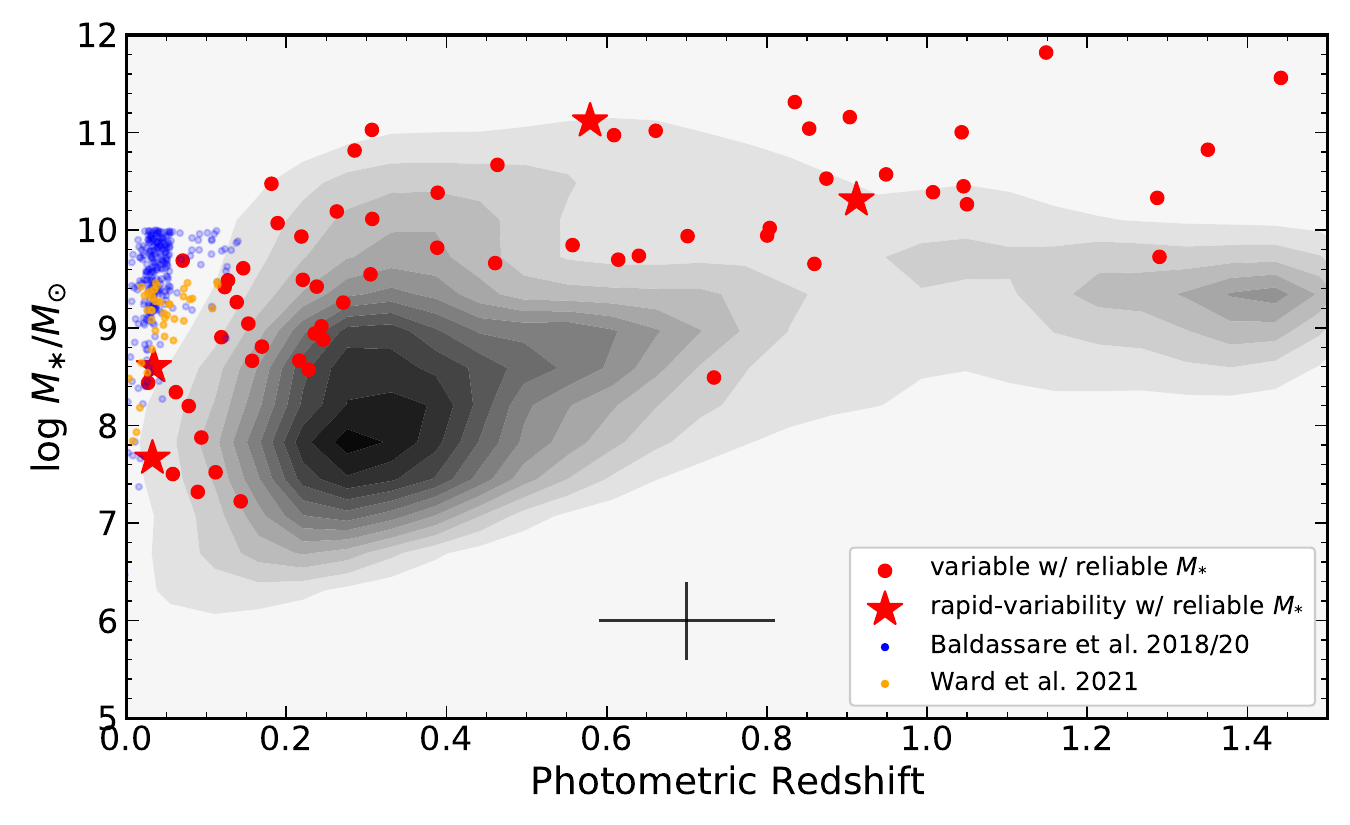}
}
\caption{Stellar mass versus photometric redshift for variable (red symbols) and non-variable (black contours) galaxies in our DES deep field sample with ``reliable'' stellar mass estimates (see \S~\ref{sec:stellarmass} for details). The sources with rapid optical variability in Table~\ref{tab:catalogvar} are shown as red star symbols. The typical (mean) uncertainties in stellar mass and redshift are shown at the bottom of the plot. For comparison, we show the variability-selected dwarf AGNs from SDSS/PTF (blue dots; \citealt{Baldassare2018,Baldassare2020}) and ZTF imaging (orange dots; \citealt{Ward2021b}). Our deep-field variability-selected AGNs are at relatively lower mass at a given redshift than previous optically-selected AGNs. This can only be matched by deep X-ray/radio imaging \citep{Guo2020}. \label{fig:redshift}}
\end{figure*}

We use the \textsc{cigale} code \citep{Burgarella2005,Noll2009,Boquien2019} to estimate the stellar masses by fitting the broadband spectral energy distribution (SED) composed of deblended stacked, model-based $ugrizJHK_{S}$ photometry. We performed SED fitting with both resolved and unresolved sources. \textsc{cigale} works by imposing a self-consistent energy balance constraint between different emission and absorption mechanisms across the EM spectrum. A large grid of models is computed and fitted to the data, allowing for an estimation of the star formation rate, stellar mass, and AGN contribution via a Bayesian-like analysis of the likelihood distribution. We use the $z_{\rm{ph}}$ values determined in \S\ref{sec:photoz} as input to \textsc{cigale} because \textsc{cigale} is not designed for $z_{\rm{ph}}$ inference, and indeed we found that the \textsc{cigale} photometric redshifts are much worse than the $z_{\rm{ph}}$ values using the method of \citet{Yang2017}. We caution that the resulting stellar mass uncertainties do not include the additional uncertainty from the covariance between redshift and stellar mass. However, the systematic uncertainties due to model choices typically dominate \citep{Ciesla2015,Boquien2019}. In addition, variability contributes additional uncertainty in the SED shape over the quoted deep field photometry given the non-simultaneity of the observations between bands. Therefore, we sum the RMS variation of the \emph{g}-band DES light curve in quadrature to the quoted photometric errors in the deep field photometric catalog in all bands.

We use a delayed exponential star formation history and vary the $e$-folding time and age of the stellar population assuming a sub-solar metallicity (dwarf galaxies are expected to follow the mass metallicity relation; \citealt{Kirby2013}.) Studies of local group galaxies have shown that a delayed exponential model of star formation history is a good approximation for dwarf galaxies \citep{Weisz2014}. If a dwarf galaxy does not undergo merger induced burst of star formation, a single delayed exponential model may be reasonable. Furthermore, \citet{Zou2022} found that differences in star formation history result in systematic differences in stellar mass of only $\sim 0.1$ dex for a sample of $z = 0 - 6$ AGN. We adopt the commonly-used \citet{Chabrier2003} initial stellar mass function with the stellar population models of \citet{Bruzual2003}. We adopt the nebular emission template of \citet{Inoue2011}. We use the \citet{Leitherer2002} extension of the \citet{Calzetti2000} model for reddening due to dust extinction, and the \citet{Draine2014} updates to the \citet{Draine2007} model for dust emission. Finally, we adopt the SKIRTOR clumpy two-phase torus AGN emission model \citep{Stalevski2012,Stalevski2016} allowing for additional polar extinction. We assume Type-1-like inclination angle varying from $i=10$ to $30$ deg for galaxies with variable light curves (\S\ref{sec:selection}), while allowing the inclination angle to vary between $i=40$ to $90$ deg for non-variable galaxies. Our choice of considering only a few viewing angles close to the average values for Type I and II AGNs is justified by previous studies, which found that different viewing angles were largely degenerate with the average values of 30 and 70 degrees for Type I and II AGNs, respectively \citep[e.g.,][]{Mountrichas2021,Padilla2021}.

There is a strong degeneracy between blue colors from AGN UV continuum emission and ongoing star-formation in the SED of an AGN plus host galaxy. This presents considerable challenges when trying to estimate a stellar mass for our variable sample if the intrinsic AGN emission is mistakenly fit as the star-formation component in \textsc{cigale}. To address this challenge, we fit each of our variable AGNs twice using \textsc{cigale}. First, instead of fixing the AGN fraction ($f_{\rm{AGN}}$) at $1~\mu$m, we float this parameter to model star-formation plus AGN, and we impose a bound of $0.2 < f_{\rm{AGN}} < 0.95$. Second, we set $f_{\rm{AGN}}=0.9999$ to model an AGN-dominated SED \citep{Yang2022}. The lower limit of 0.2 is chosen as a conservative lower limit on the AGN luminosity fraction and is broadly consistent with variability of dwarf AGNs \citep{Burke2022}. We compute the reduced $\chi^2$ values for each, and compute the difference $\Delta \chi_{\nu}^2 = (\chi_{\nu}^2)_{f_{\rm{AGN}}=1} - \chi_{\nu}^2$. We interpret those SED with an improved fit when $f_{\rm{AGN}}$ is a free parameter to be sources with a significant contribution from star-formation emission using the criteria $\Delta \chi_{\nu}^2 > 2$. Because these sources have a significant star-formation component in their SEDs, we consider their stellar masses to be reliable. On the other hand, sources with SEDs dominated by their AGN are unlikely to have a well constrained star-formation component, and therefore the stellar mass estimates are not reliable. We reject $\sim 1$ percent sources with a poor best-fit model by requiring $\chi_\nu^2<10$. The majority of the bad fits are sources with anomalous photometry or stars mis-classified as galaxies. 

A difficult problem is the degeneracy between low-redshift starburst galaxies and high-redshift quasars with blue colors from AGN UV continuum emission. Because we are interested in selecting variable dwarf galaxies, quasars incorrectly identified as low-redshift star-forming galaxies (which tend to be low-mass) are a major contaminant. Figure~\ref{fig:resolved_frac} demonstrates the effect of quasars with high-redshifts incorrectly identified as dwarf galaxies. This branch of failures primarily occurs with high-redshift sources with incorrect photometric redshifts near $z_{\rm{ph}} \sim 0.1-0.4$, because of a color degeneracy at these redshifts. These quasars have AGN power-law dominated emission that are mostly rejected by our $\Delta \chi_{\nu}^2 > 2$ constraint.

After this, the final parent sample of galaxies with well-sampled light curves and acceptable SED fits is \Nparent{}. Of these, we find \Nvar{} variable galaxies, and \Ndwarf{} have $M_{\ast}<10^{9.5}\ M_{\odot}$ with reliable stellar mass estimates. The resulting distribution of stellar mass and redshift is shown in Figure~\ref{fig:redshift}. To validate our SED fitting results, we compare our results in SN-C3 to matched sources with stellar masses and photometric redshifts from the FourStar Galaxy Evolution Survey (ZFOURGE; \citealt{Tomczak2014,Straatman2016}) in Appendix~\ref{sec:stellarmassvalidation}. One concern is AGNs that rejected using our AGN-dominated model comparison could be a function of stellar mass if the AGN-dominated source is more massive/less star-forming. This bias may impact the variability fraction at larger stellar masses. Finally, we caution that additional systematic uncertainties on the stellar mass may be up to 20 percent due to uncertainties in stellar evolution (e.g., initial mass function, star formation history) even when the photometric redshift is accurate \citep{Ciesla2015,Boquien2019}. This does not include additional sources of error from deneracies between star-formation and AGN light, for instance. Nevertheless, we check our photometric redshifts against spectroscopic redshifts available from the literature. The results are shown in Appendix~\ref{sec:stellarmassvalidation}.

\subsection{Variability Analysis}

\begin{figure}
\includegraphics[width=0.48\textwidth]{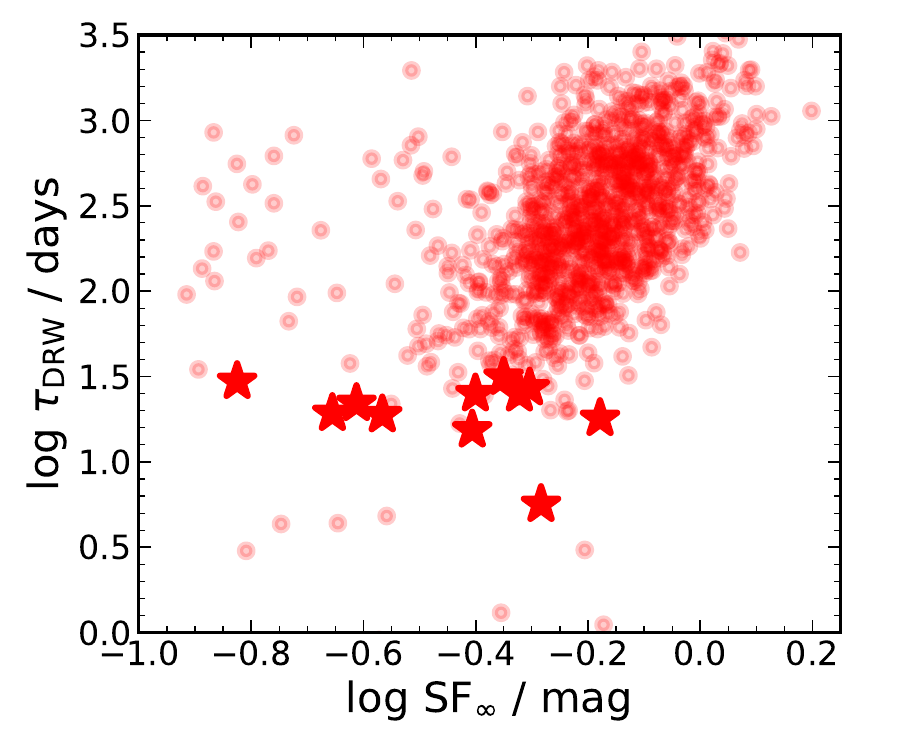}
\centering
\caption{Rest-frame damping timescale $\tau_{\rm{DRW}}$ versus asymptotic variability amplitude ${\rm{SF}}_{\infty}$ for our AGN candidates. The sources with rapid optical variability with constrained damping timescales in Table~\ref{tab:catalogvar} are shown as red star symbols. \label{fig:SFtau}}
\end{figure}

To investigate the variability properties of our AGN candidates, we follow the Bayesian method of \citet{Kelly2009} using a damped random walk (DRW) prescription. The DRW is a common model which can describe the stochastic fluctuations in AGN optical light curves, which may result from thermal fluctuations in the accretion disk. The simple DRW model includes both an amplitude $\rm{SF_{\infty}}$ and characteristic (damping) variability timescale term $\tau_{\rm{DRW}}$.  We use the Gaussian process DRW prescription following \citet{Burke2021}, which makes use of the \textsc{celerite} \citep{Foreman-Mackey2017} and \textsc{emcee} \citep{Foreman-Mackey2013} codes. The amplitude$-$rest-frame timescale distribution of our sources is shown in Figure~\ref{fig:SFtau}. 

Recently, \citet{Burke2021} identified a scaling relation between the characteristic timescale and BH mass using AGN light curves. This enables a BH mass estimate independent of spectroscopic techniques or via indirect stellar mass estimation using SED fitting, which can suffer from strong model degeneracies or rely on the BH--host scaling relations which are still poorly constrained in dwarf galaxies particularly at high redshift. The relation is given by,
\begin{equation}\label{eqn:tau_mass}
    M_{\rm{BH}} = 10^{7.97^{+0.14}_{-0.14}}\ M_{\odot}\   \left(\frac{\tau_{\rm{DRW}}}{100\ \rm{days}}\right)^{2.54^{+0.34}_{-0.35}}\ ,
\end{equation}
with an intrinsic scatter of $0.33\pm0.11$ dex in $M_{\rm{BH}}$. Therefore, a reasonable variability constraint for dwarf AGN identification is $\log( \tau_{\rm{DRW}} {\rm{/ days}})\leq1.5$, which corresponds to $M_{\rm{BH}} \lesssim 5\times10^6\ M_{\odot}$. We also impose a requirement that the observed-frame damping timescale be larger than the observed cadence to avoid unconstrained values. Finally, we caution that the damping timescales for some of our sources may be biased smaller to due insufficient light curve duration \citep{Kozlowski2017}. This is unlikely to affect our dwarf AGNs, whose damping timescales are typically less than ten times the light curve baseline. Finally, we note this relation has some scatter that we are selecting against which can result in a large scatter in the resulting stellar masses, even if they are reliable due to the $0.5$ dex scatter in the host galaxy-BH mass scaling relations \citep{Reines2015}. One source, J033051.6$-$272856.2 has an apparently anomalously large stellar mass ($M_\ast \sim 10^{11}\ M_\odot$) but its SED has a large contribution from AGN emission which could bias the stellar mass larger (see Figure~\ref{fig:examples}.)


\section{Results} \label{sec:results}

\begin{figure}
\centering
\includegraphics[width=0.5\textwidth]{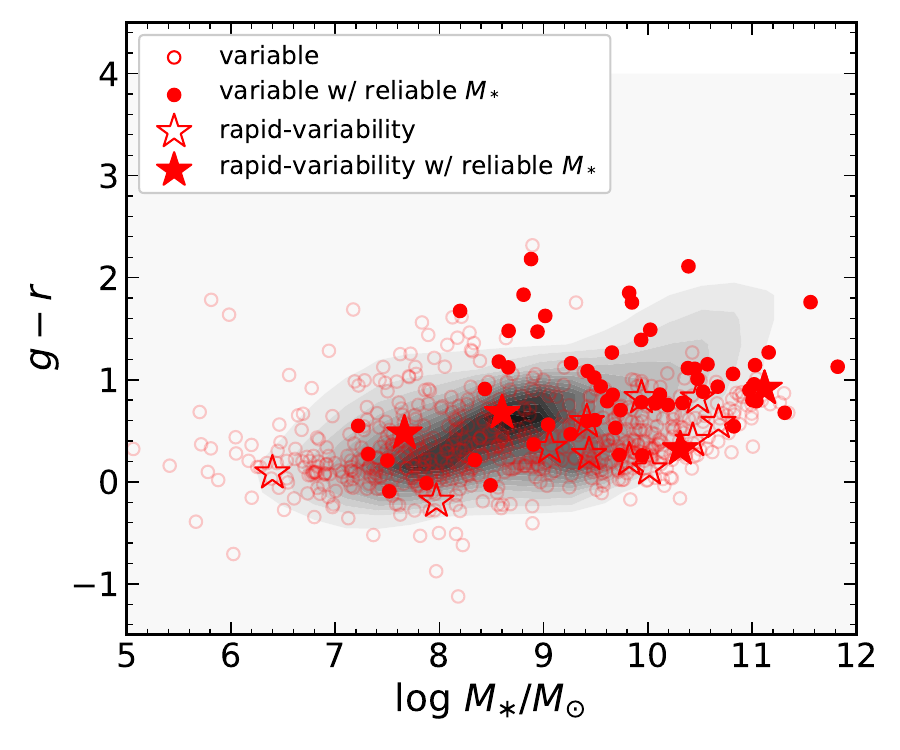}
\caption{Galaxy \emph{g-r} colors versus stellar mass for variable (red) and non-variable (black contours) sources. The variable galaxies tend to be somewhat bluer at given stellar mass. \label{fig:colors}}
\end{figure}

\subsection{Detection Fraction} \label{sec:detfrac}

\begin{figure*}
{
\includegraphics[width=1\textwidth]{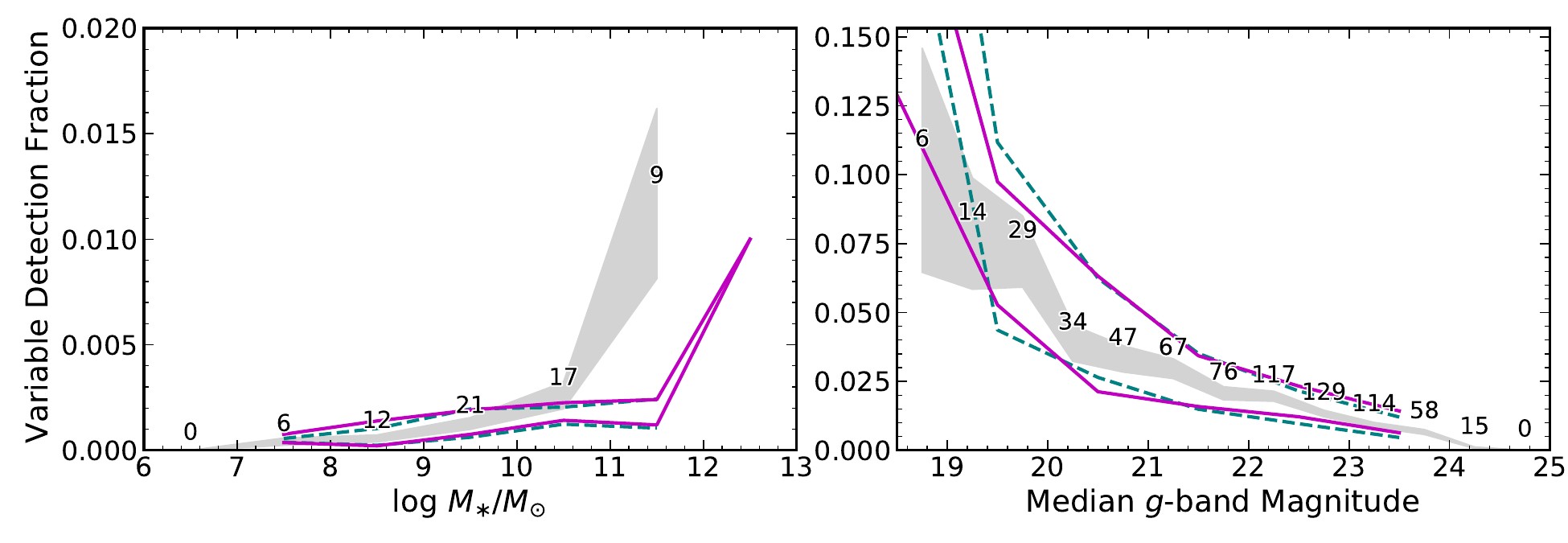}
}
\caption{Fraction of variable AGN versus stellar mass (\emph{left panel}) and median \emph{g}-band apparent magnitude (\emph{right panel}). The gray shaded areas are the $1\sigma$ bands of uncertainty for each bin computed assuming a binomial distribution. The number of variable galaxies in each bin is given above each bin. The magenta lines are the predicted detection fractions assuming a model with a constant occupation fraction of 1 (the most optimistic ``light'' seed scenario). The green dashed lines are the predictions assuming an occupation fraction which drops dramatically below $M_*=10^8\ M_{\odot}$ (the ``heavy'' seed scenario; \citealt{Ricarte2018,Bellovary2019}), as described by \citet{Burke2022} for central BHs (see \S\ref{sec:detfrac} for details). We caution that the bright end with $g \lesssim 18$ mag and $M_{\ast} \gtrsim 10^{10.5}\ M_{\odot}$ is highly incomplete and our model and observations differ. \label{fig:detfrac}}
\end{figure*}

We select \Nvar{} variable AGN candidates out of \Nparent{} total galaxies. We find \Ndwarf{} AGNs in low stellar mass $M_\ast<10^{9.5}\ M_\odot$ galaxies with reliable stellar mass estimates. We show their optical colors versus stellar mass in Figure~\ref{fig:colors}. The stellar mass versus redshift of our variability-selected candidates are shown in Figure~\ref{fig:redshift}. We plot the fraction of variability-selected AGNs versus magnitude and stellar mass in Figure~\ref{fig:detfrac}. However, the variable AGN detection fraction is influenced by a selection bias of variability being more difficult to detect in fainter sources. However, an understanding of the constraints on the occupation fraction from the observed AGN variability fraction requires a comprehensive demographic model of the true variable AGN population combined with physically-motivated AGN light curve simulations to capture the selection effects related to the survey sensitivity, depth, and light curve sampling.

We attempt to quantify these effects using the forward Monte Carlo sampling model of \citet{Burke2022}. This model generates mock light curves for a given instrument from a population of variable AGNs drawn from ``light'' and ``heavy'' seeding scenarios (e.g., \citealt{Ricarte2018,Bellovary2019}) using an input galaxy stellar mass function, Eddington ratio distribution function, obscured AGN fraction, and constrains on AGN variability behavior from observations. We input the DES-SN-like survey parameters (6 year baseline, 7 day cadence) using the typical photometric precision shown in Appendix~\ref{sec:errorbarcorrection}. We do not include off-nuclear variable black holes in our comparison because our study is restricted to nuclear variability. It is unclear how the off-nuclear IMBH population, which could make up a larger fraction of AGNs in dwarf galaxies, could be connected to their host galaxy stellar mass (e.g., \citealt{Greene2020}) A stellar mass uncertainty of 0.6 dex is assumed in our model prediction. The distinguishing power between the two occupation fractions lies in the shapes of the variability fractions in Fig~\ref{fig:detfrac}. We re-normalized the detection fractions by an arbitrary scaling to match our detection fraction, because we have removed a large fraction of variable sources with unconstrained stellar mass estimates. Our variable fraction is lower than some previous works \citep{Baldassare2020} but more consistent with \citet{Baldassare2018}. This is, in part, dependent on the limiting redshift of the parent sample. Given the limited number of variable sources with reliable stellar masses in the $4.6$ deg$^2$ in this work, our model predictions are unable to distinguish between the two occupation fractions for the different ``light'' and ``heavy'' seeding scenarios. Hence, we are unable to put strong constraints on the occupation fraction at this time. Future work including a larger area with greater number statistics may be more promising. Nevertheless, we have demonstrated the feasibility of using deep fields to explore the variable dwarf AGN population.

\subsection{Dwarf AGN Candidates}

\begin{figure*}
\centering
\begin{minipage}{.5\textwidth}
\centering
\includegraphics[width=0.8\textwidth]{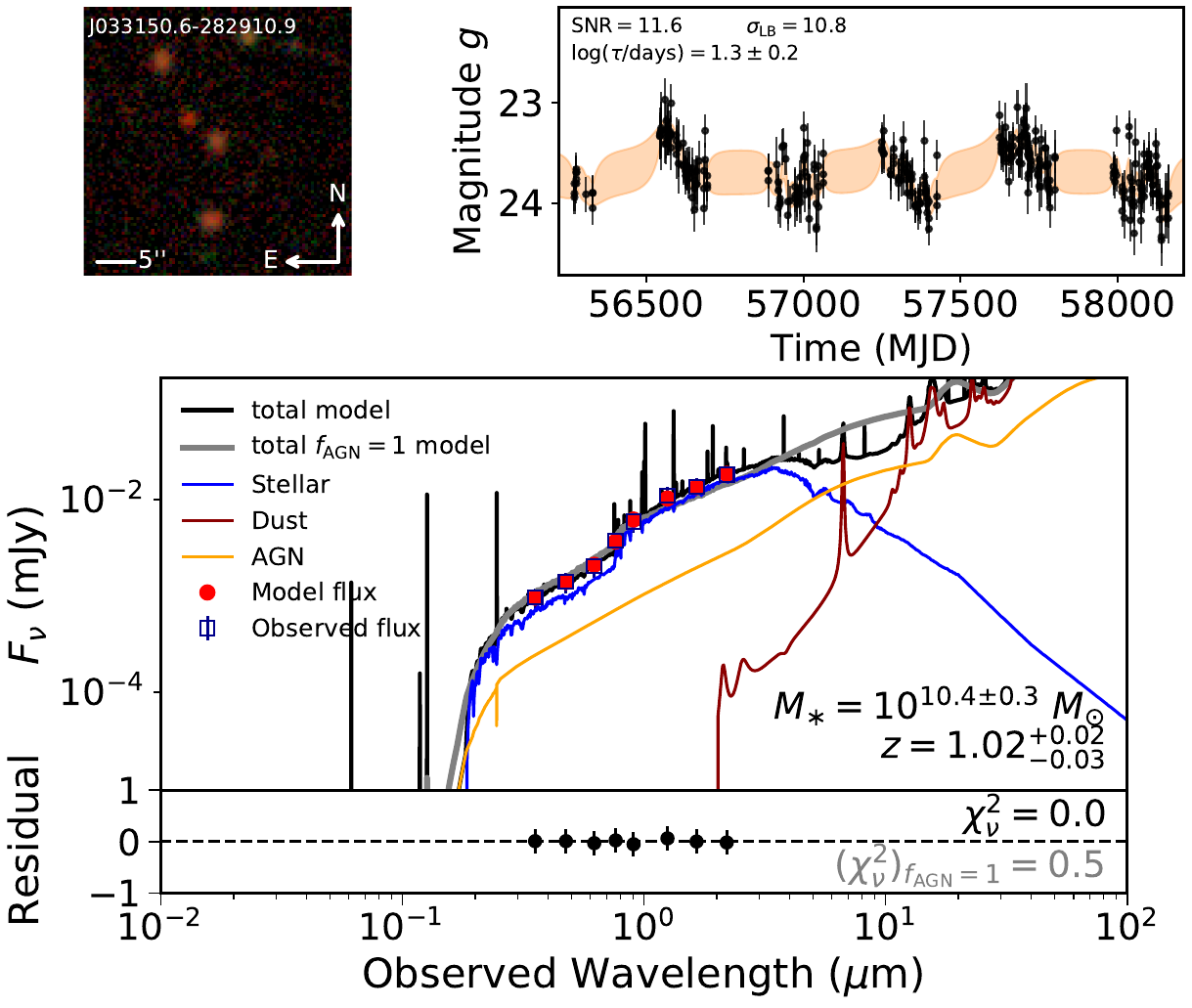}
\includegraphics[width=0.8\textwidth]{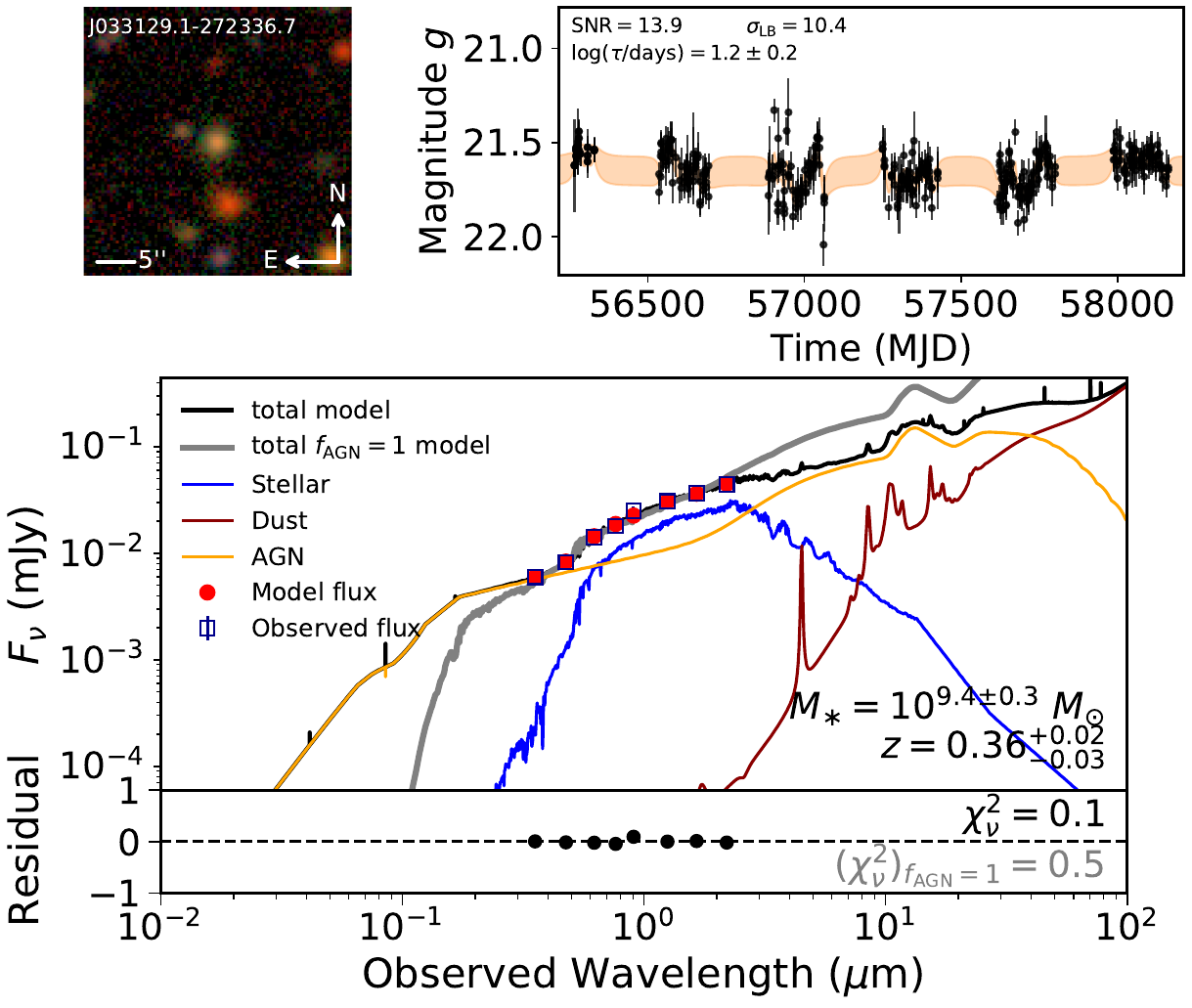}
\includegraphics[width=0.8\textwidth]{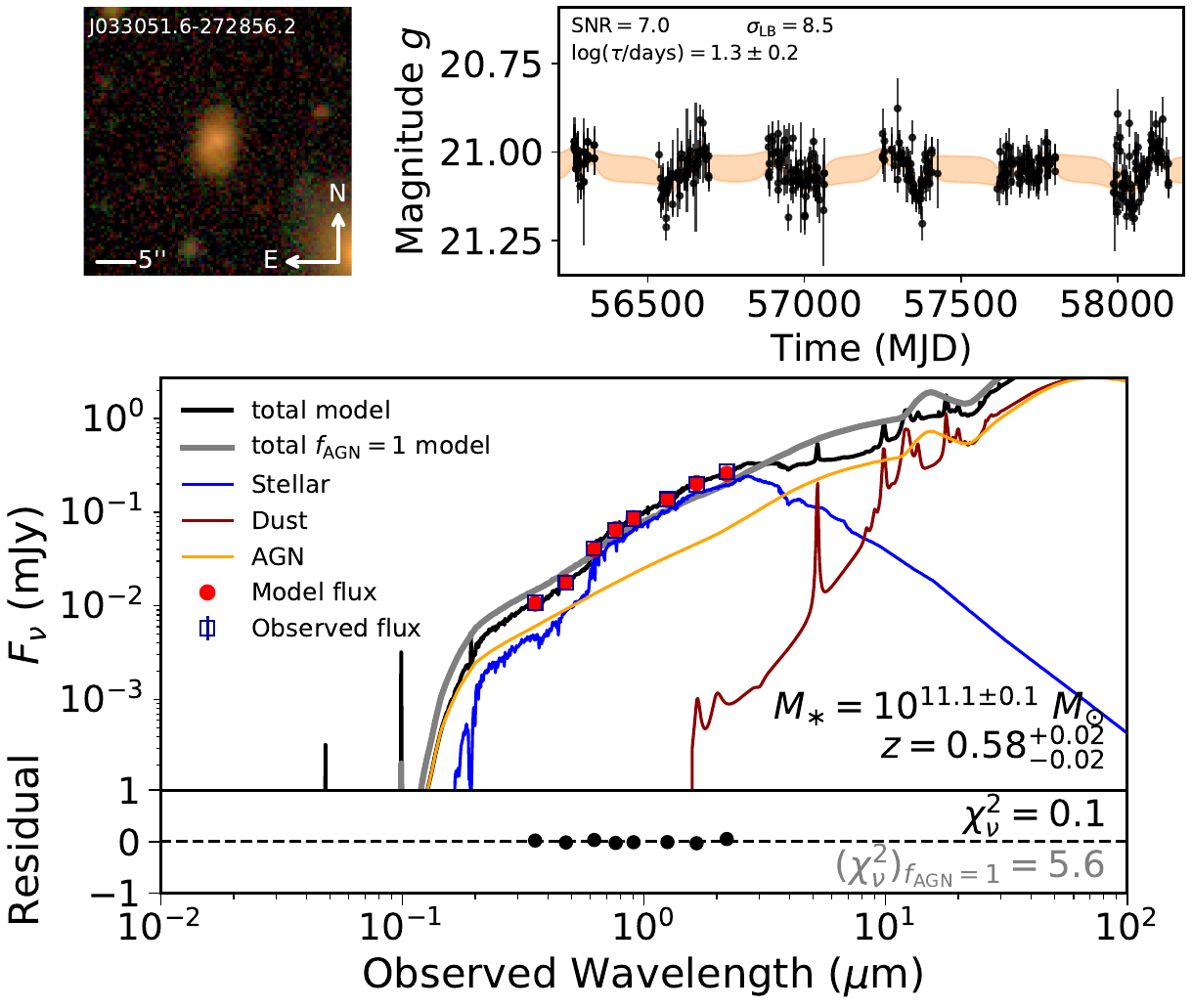}
\end{minipage}%
\begin{minipage}{.5\textwidth}
\centering
\includegraphics[width=0.8\textwidth]{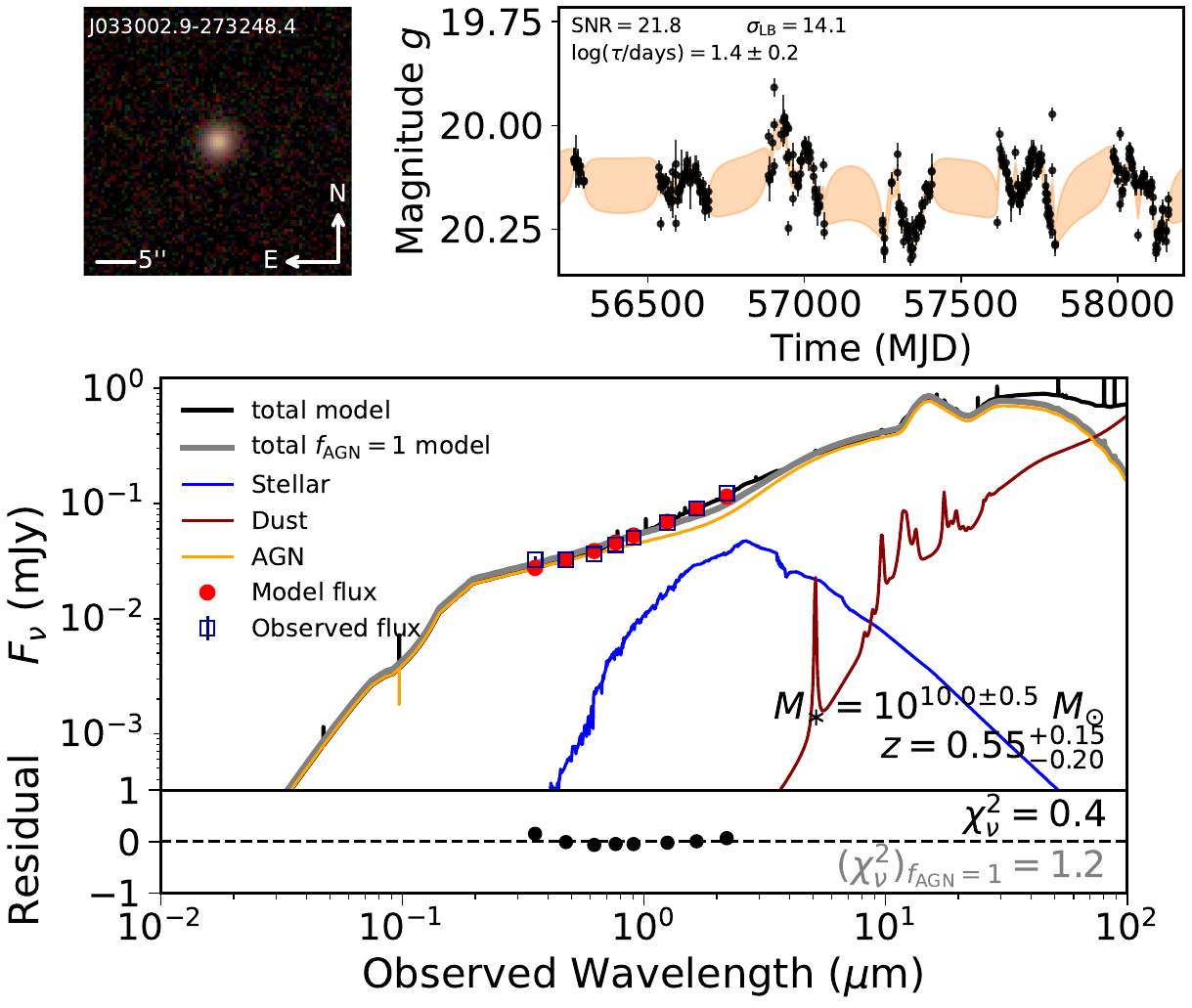}
\includegraphics[width=0.8\textwidth]{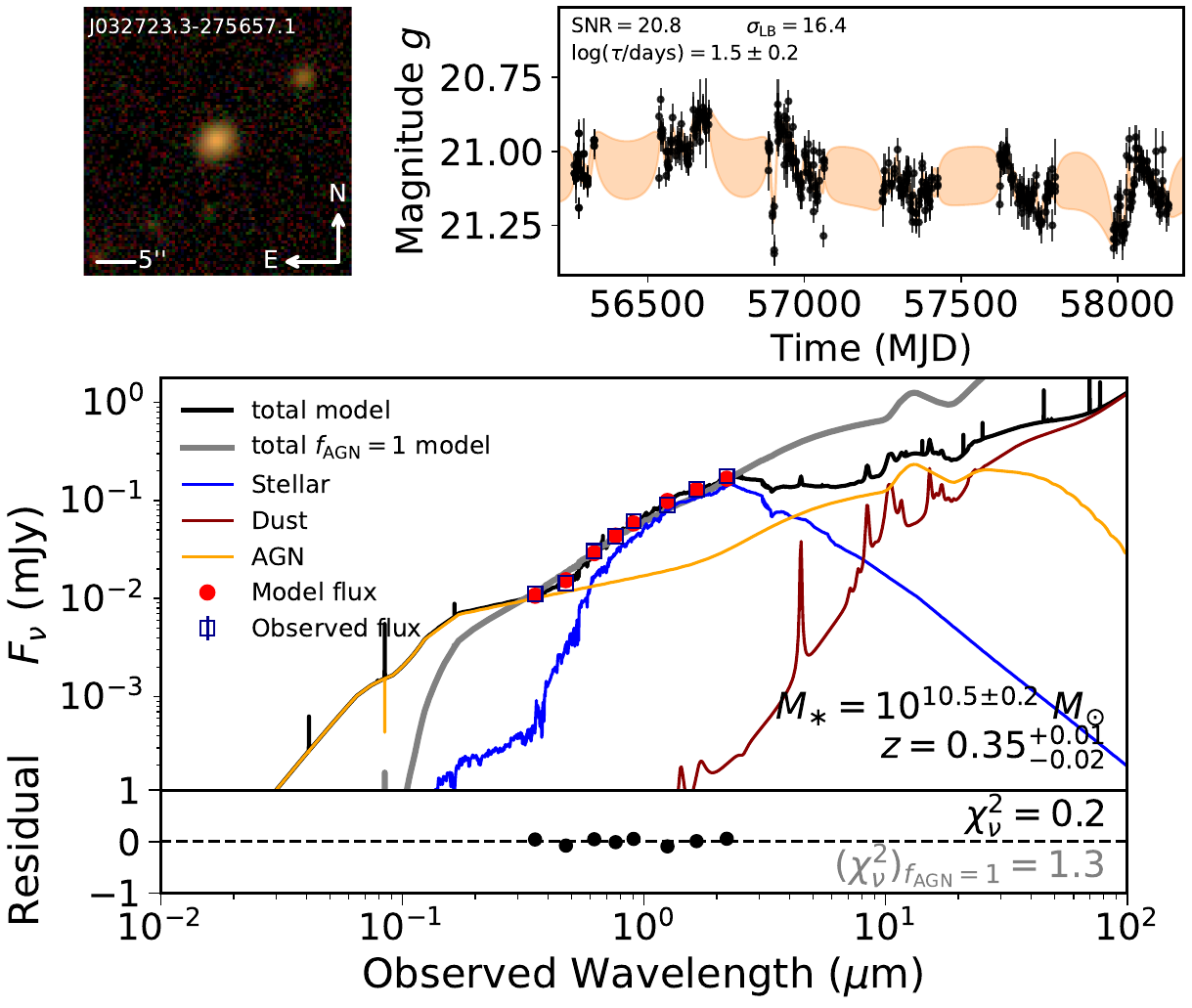}
\includegraphics[width=0.8\textwidth]{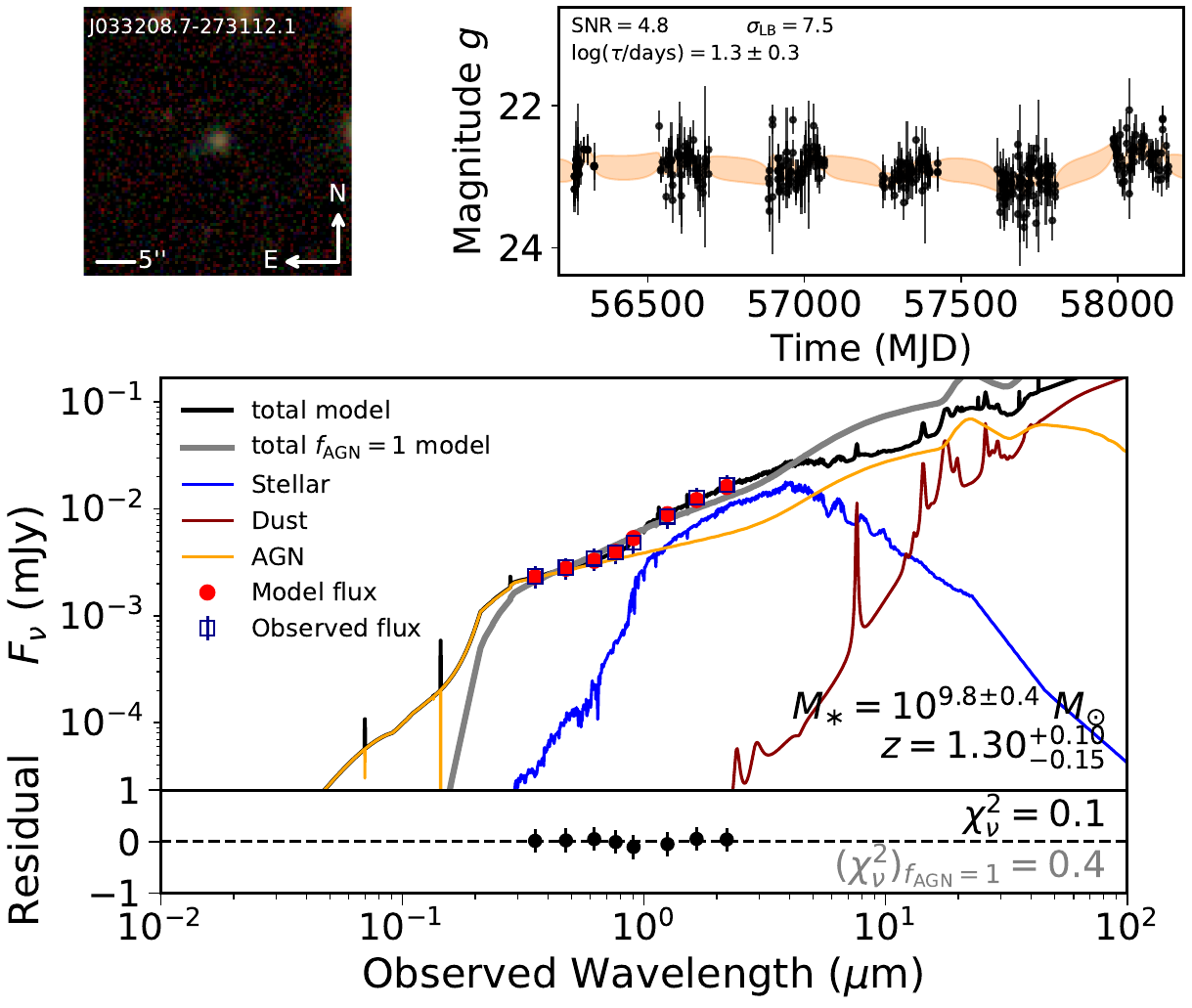}
\end{minipage}%
\caption{Candidate variability-selected dwarf AGNs from Table~\ref{tab:catalogvar}. We show the DES \emph{gri} color-composite coadd (\emph{top left}). The coadd image is stacked to the 6 year DES wide field depth. We also show the \emph{g}-band light curve as black points with the $1\sigma$ error ellipse from the DRW modeling in orange (\emph{top right}) and best-fit \textsc{cigale} SED fitting results (\emph{bottom}). The observed photometry is shown as blue squares. The best-fit model photometry is shown as red points. The best-fit AGN$+$star-forming model (allowing $f_{\rm{AGN}}$ to be a free parameter) is shown in black, while the best-fit AGN-dominated model (fixing $f_{\rm{AGN}}=1$) is shown in gray. The components from attenuated stellar emission (blue), dust emission (red), and the AGN emission (orange) are also shown. A nebular emission component is also fit, but its component is not shown for clarity. The estimated stellar mass and photometric redshift are shown in the lower-right of the panel, but the stellar mass uncertainties are likely under-estimated for reasons described in \S\ref{sec:stellarmass}. The relative residual flux [(observed $-$ model) $/$ observed] is shown at the bottom of each panel along with the $\chi^2_\nu$ for the AGN$+$star-forming model (black) and AGN-dominated model (gray). A larger $(\chi^2_\nu)_{f_{\rm{AGN}}=1}$ value indicates there is a significant star-formation component in the SED and that the stellar mass estimate is likely more reliable. \label{fig:examples}}
\end{figure*}

\begin{figure*}\ContinuedFloat
\centering
\begin{minipage}{.5\textwidth}
\centering
\includegraphics[width=0.8\textwidth]{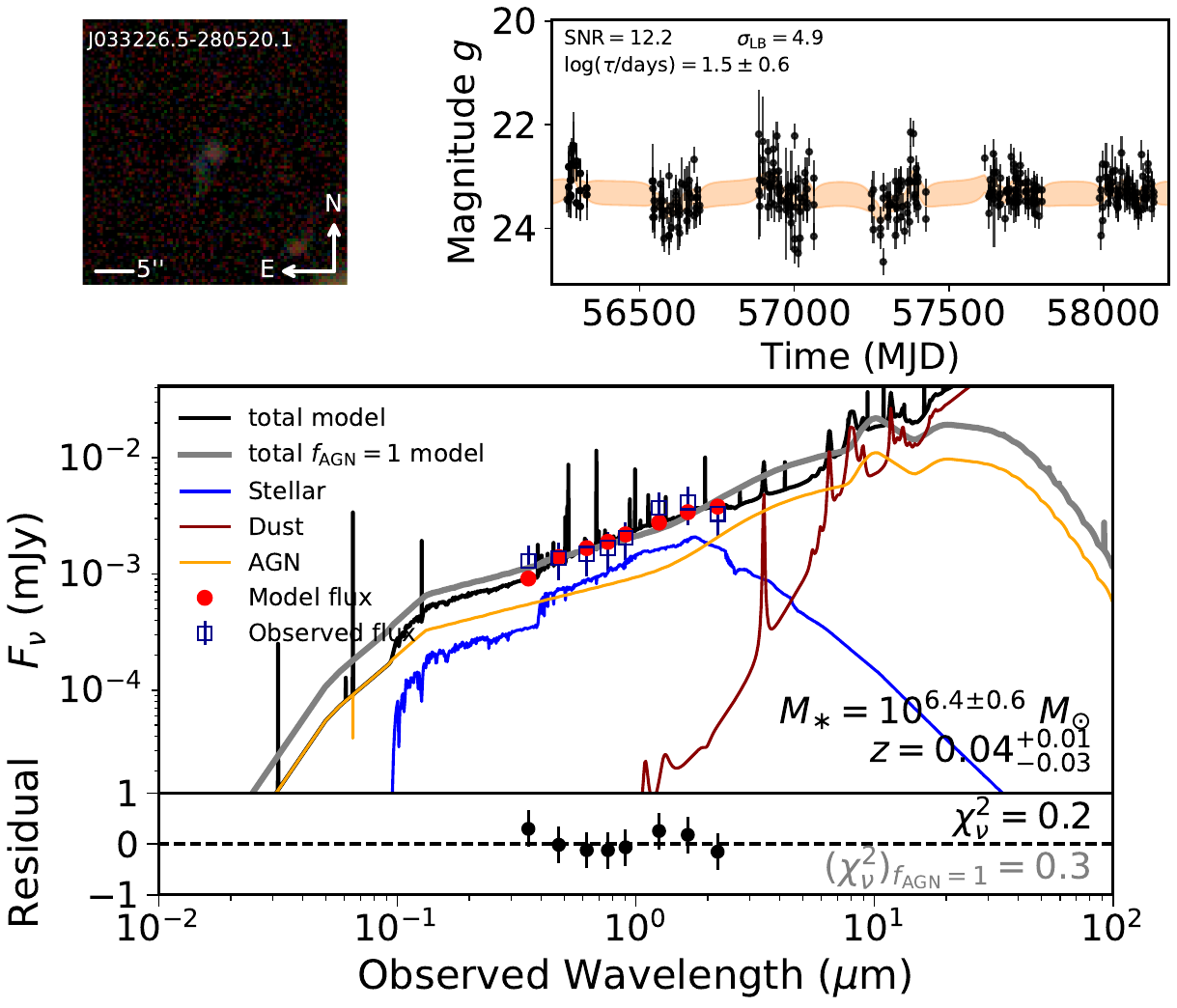}
\includegraphics[width=0.8\textwidth]{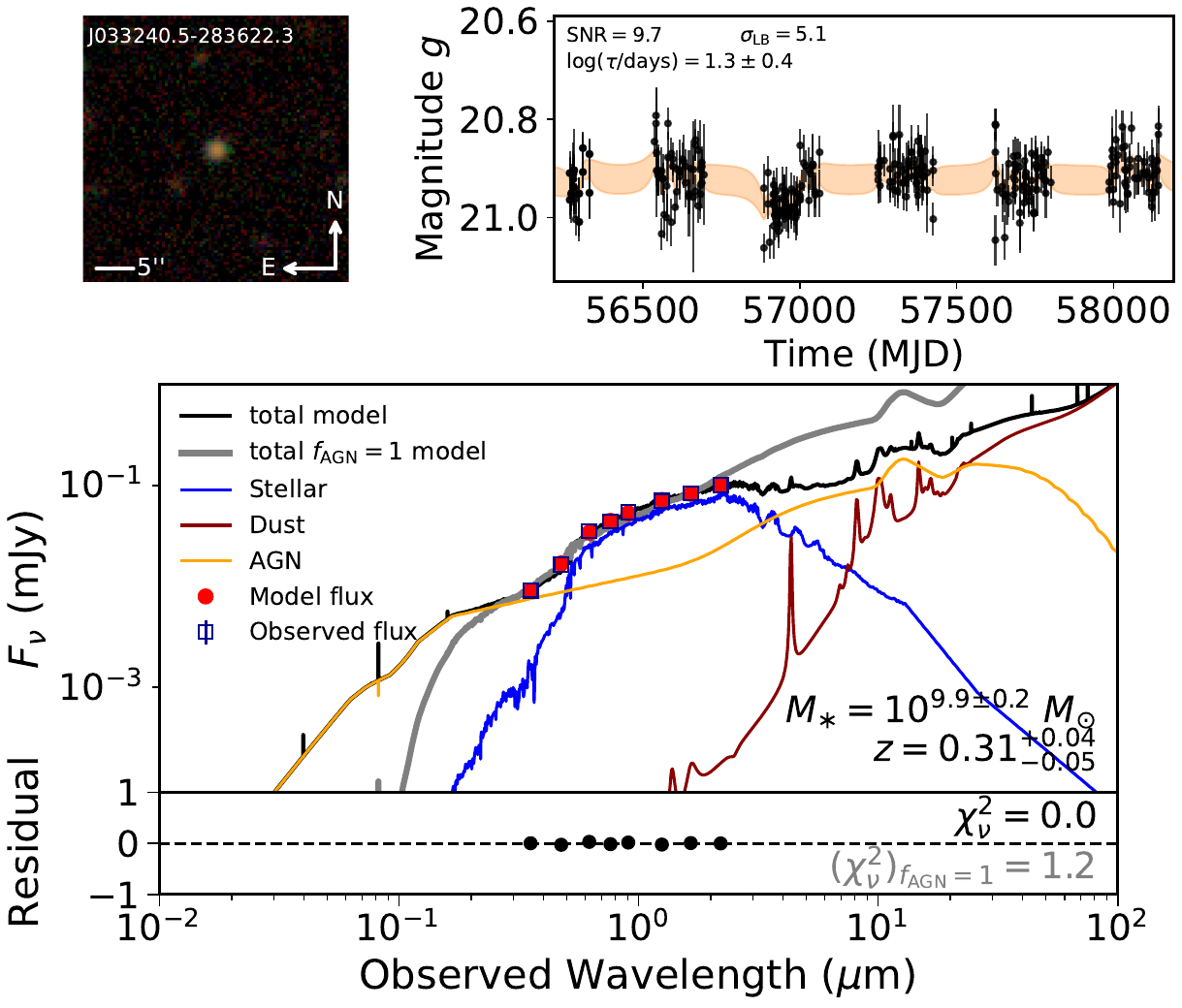}
\includegraphics[width=0.8\textwidth]{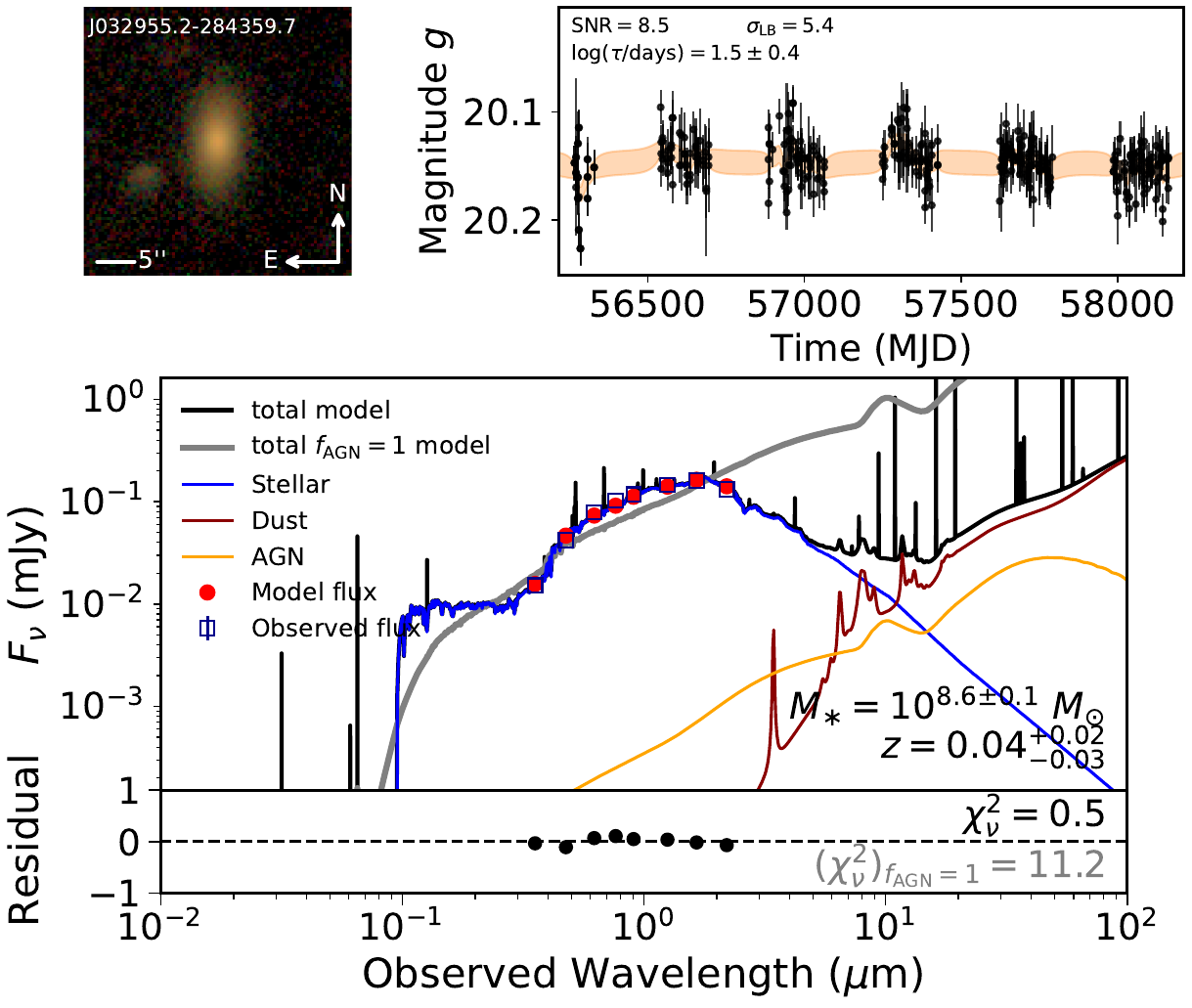}
\end{minipage}%
\begin{minipage}{.5\textwidth}
\centering
\includegraphics[width=0.8\textwidth]{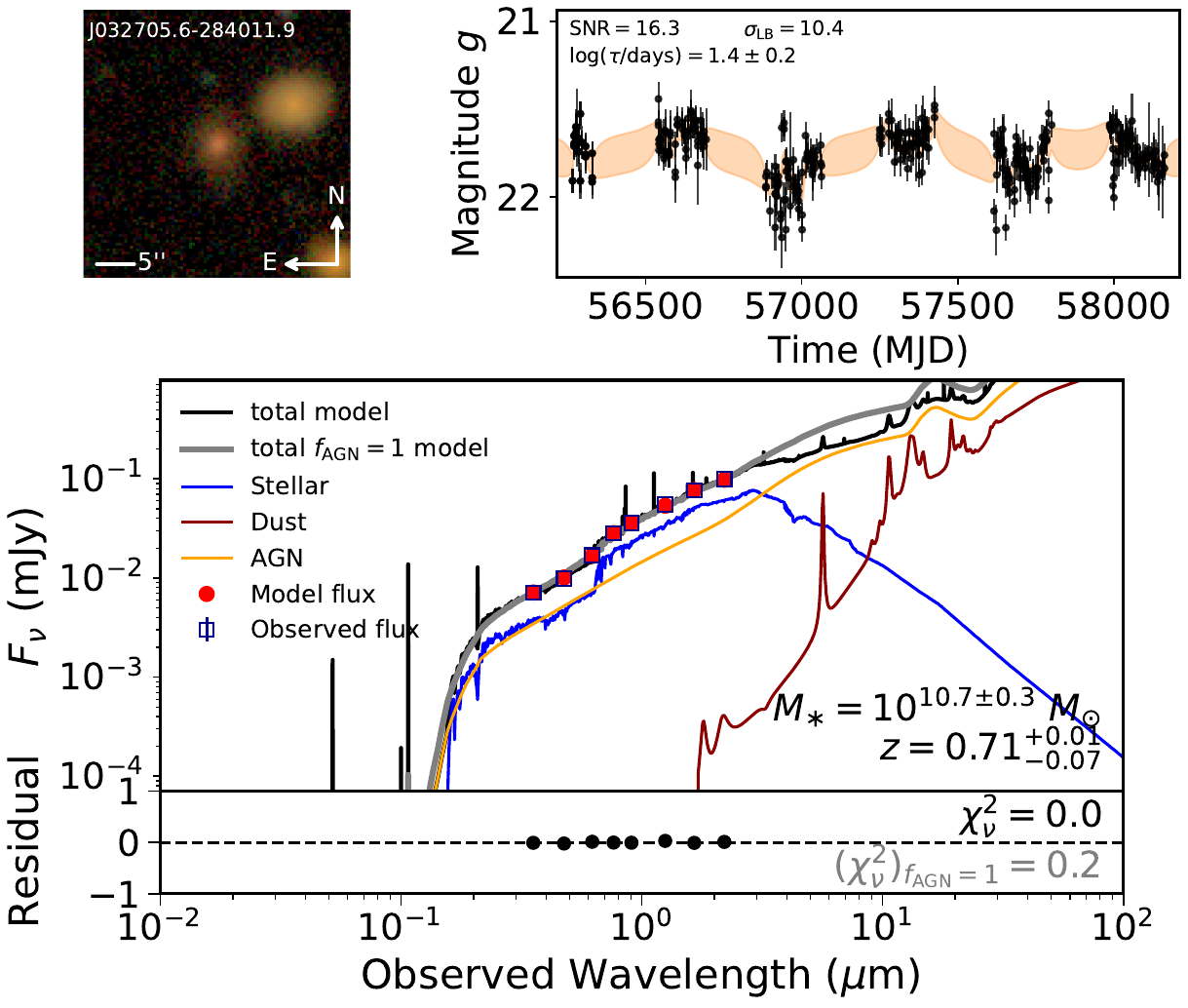}
\includegraphics[width=0.8\textwidth]{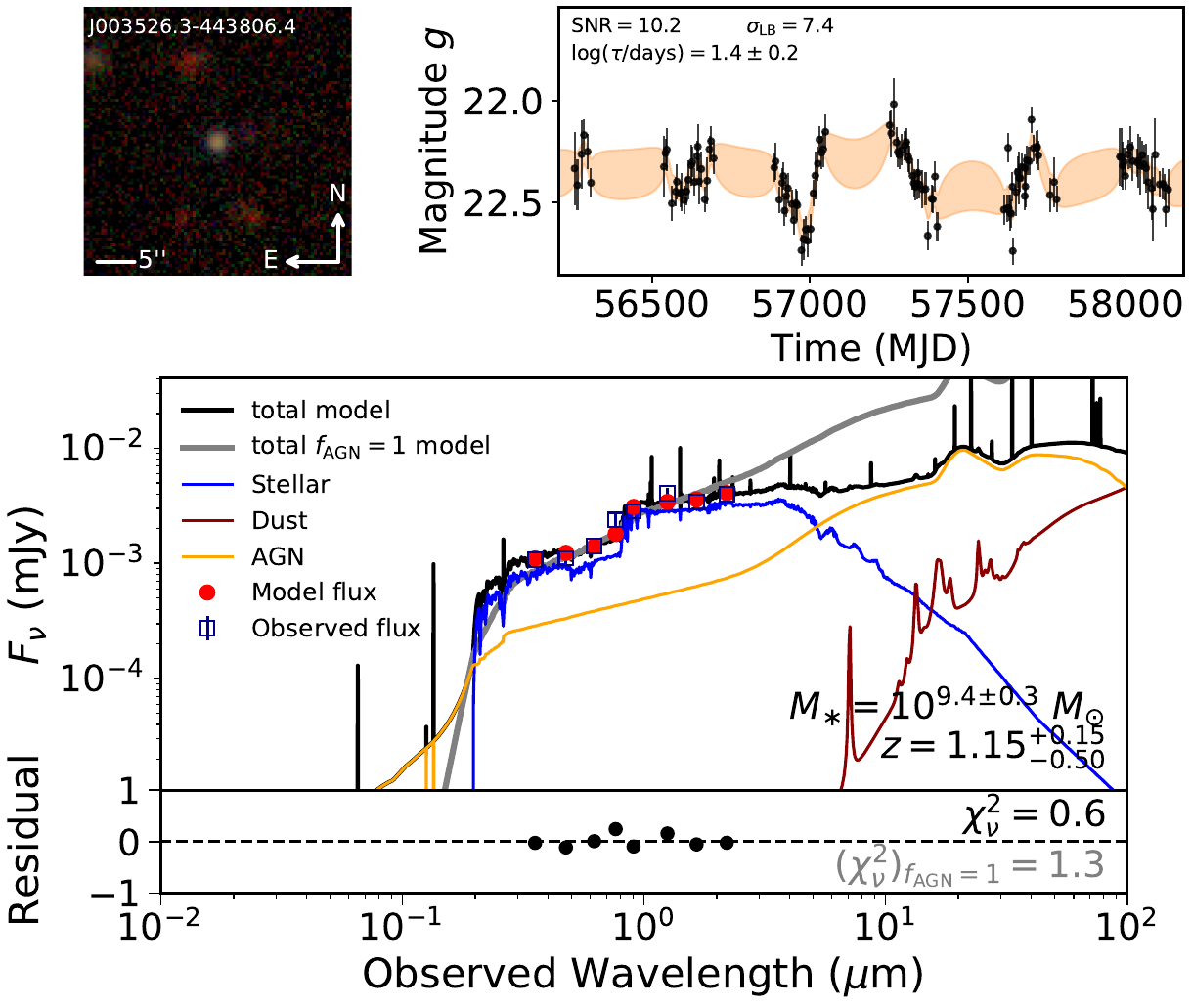}
\includegraphics[width=0.8\textwidth]{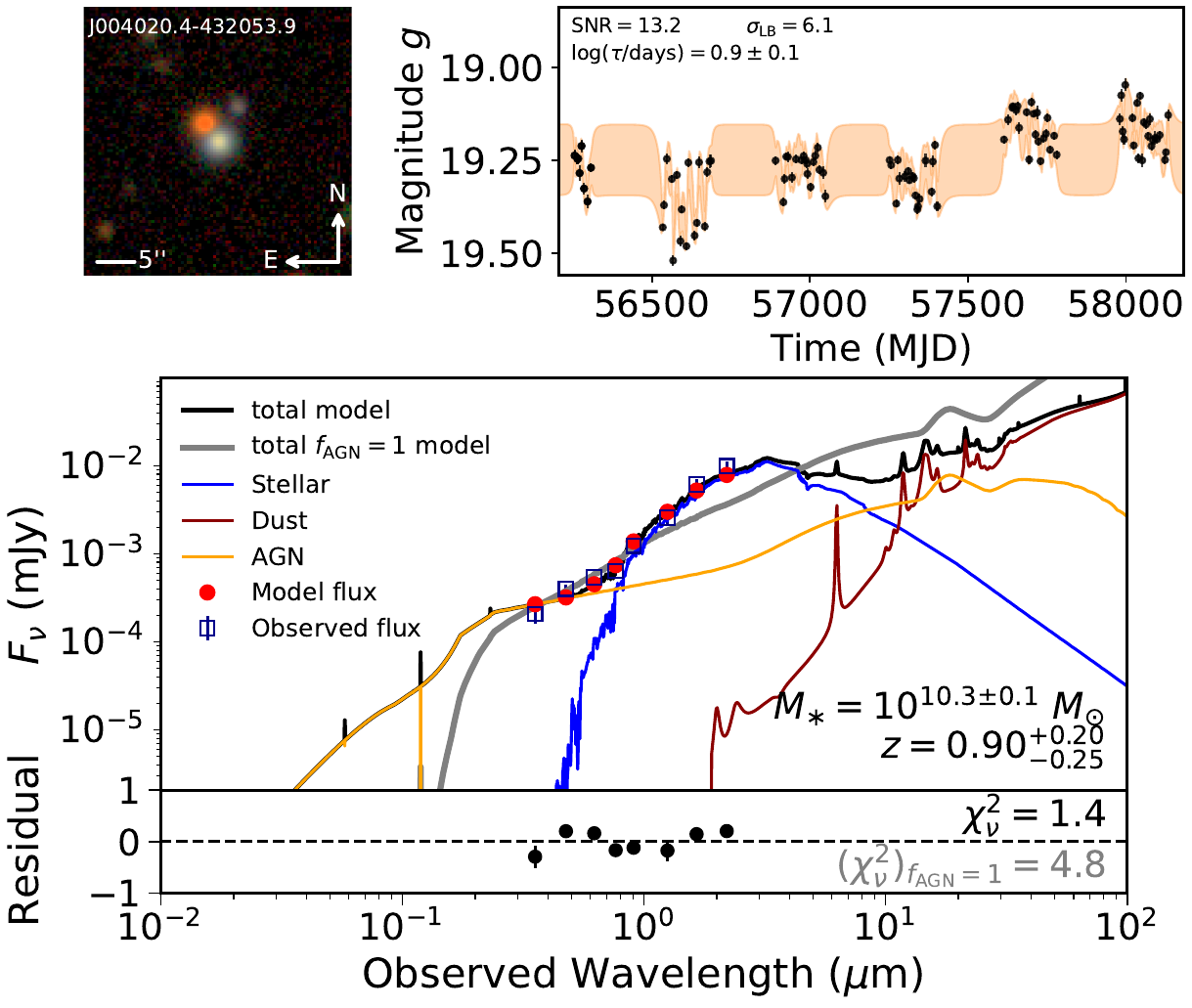}
\end{minipage}%
\caption{(Continued)}
\end{figure*}

\begin{figure*}\ContinuedFloat
\centering
\begin{minipage}{.5\textwidth}
\centering
\includegraphics[width=0.8\textwidth]{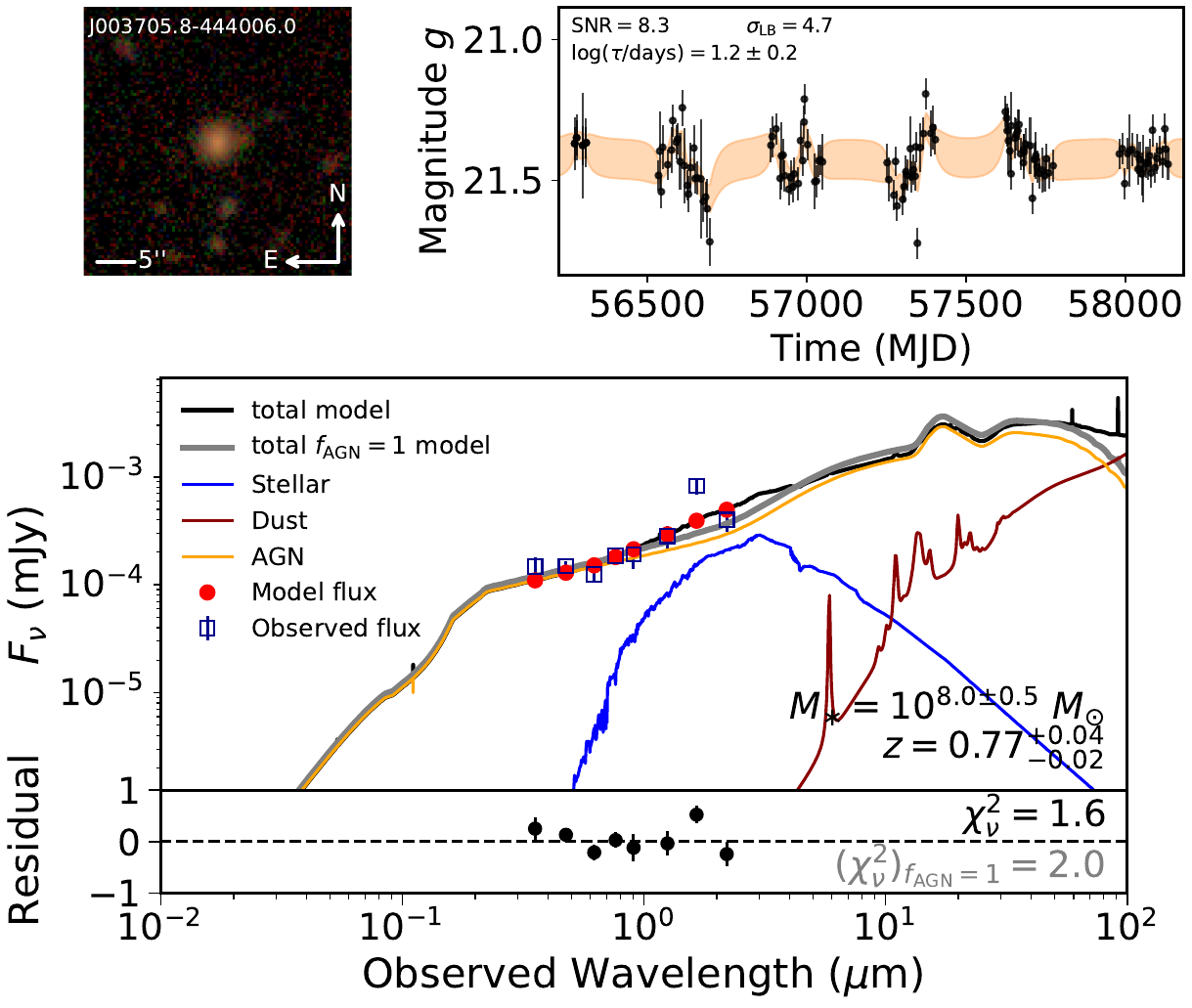}
\includegraphics[width=0.8\textwidth]{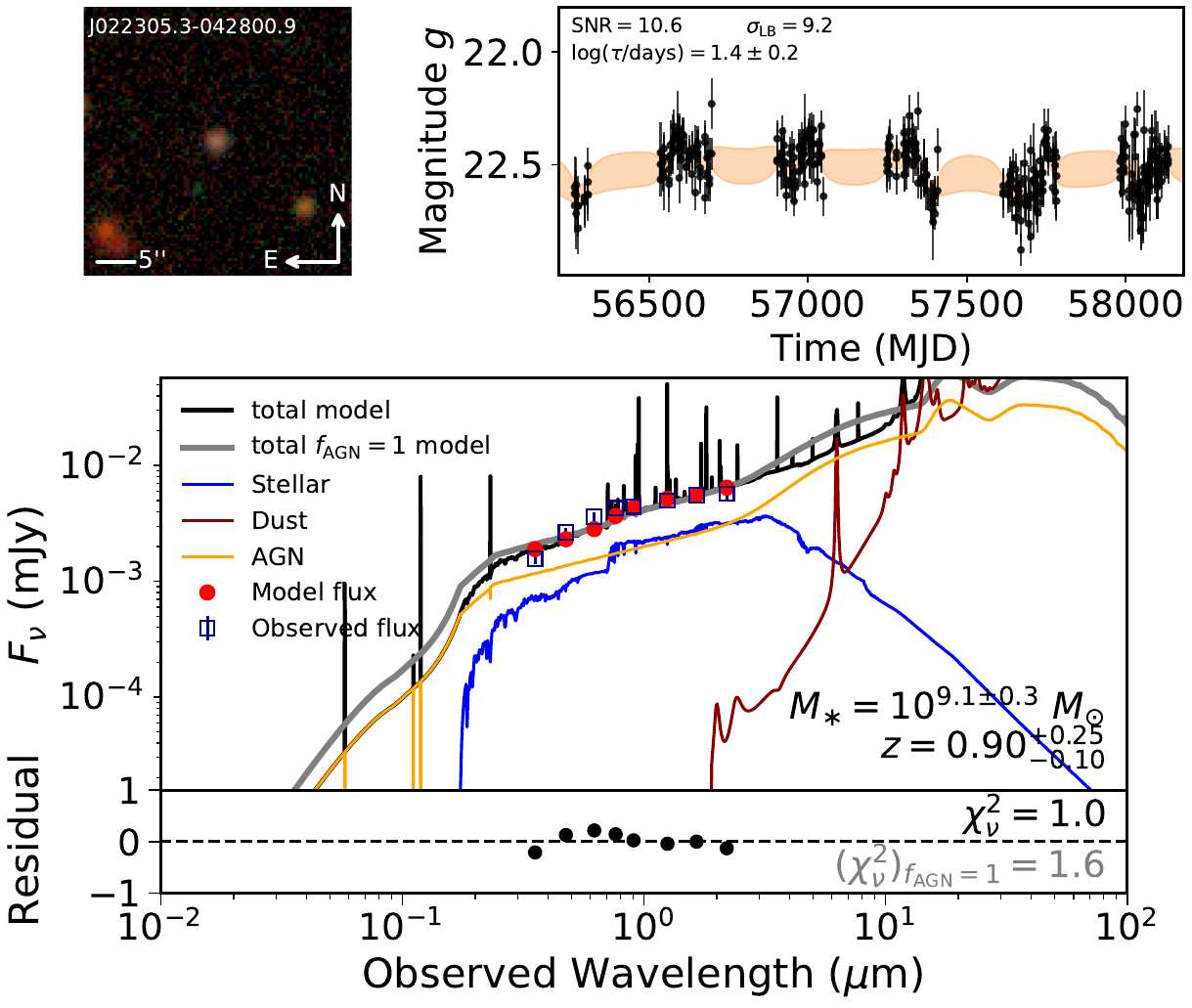}
\end{minipage}%
\begin{minipage}{.5\textwidth}
\centering
\includegraphics[width=0.8\textwidth]{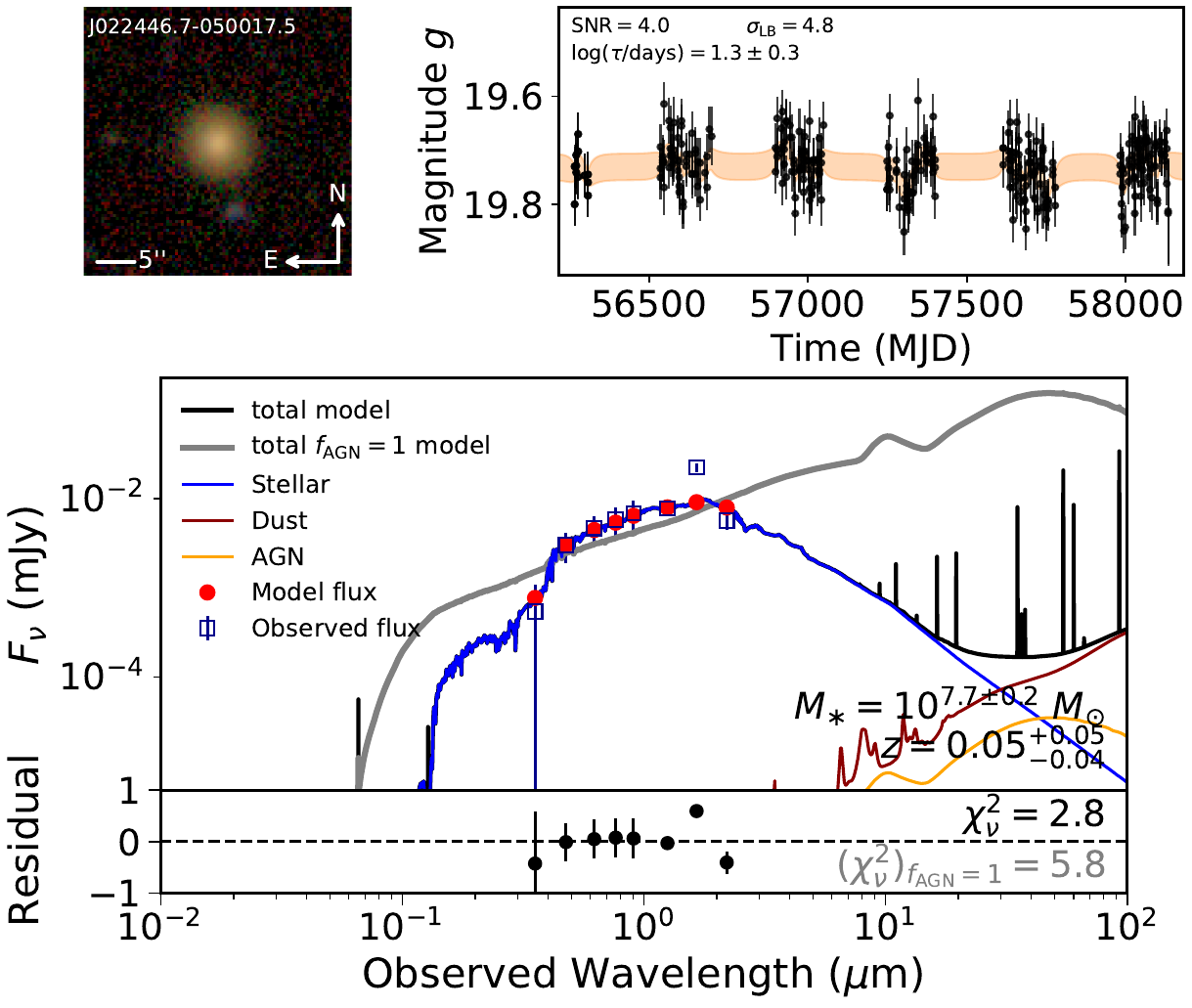}
\end{minipage}%
\caption{(Continued)}
\end{figure*}

We present a catalog of all of our variability-selected AGN candidates in Table~\ref{tab:catalogall} regardless of stellar mass or variability properties. We list the source name, coordinates, median $g$-band magnitude, variability statistics (SNR and $\sigma_{\rm{LB}}$), stellar mass estimate, and damping timescale $\tau_{\rm{DRW}}$. If available from supplementary catalogs, we include the spectroscopic redshift $z_{\rm{sp}}$ and the source of $z_{\rm{sp}}$. When available, we also include supplementary \emph{WISE} $W1-W2$ colors and the \emph{Chandra} Source Catalog hard X-ray flux. Finally, we list the DES deep field that contains the source. The stellar mass estimates should be treated with caution for reasons described in \S\ref{sec:stellarmass}. Therefore, we present a subset of this catalog in Table~\ref{tab:catalogvar}, which includes only those AGN candidates with rapid characteristic variability timescales, defined as rest-frame $\log( \tau_{\rm{DRW}} {\rm{/ days}})<1.5$ with observed-frame damping timescales greater than the 7 day cadence. Most of the \Nfinal{} sources have stellar masses below $10^{10}\ M_{\odot}$. We show our dwarf AGN candidate images, light curves, and SED fitting analysis for the candidates with rapid variability in Table~\ref{tab:catalogvar} in Figure~\ref{fig:examples}. This strict criteria is expected to result in a pure selection of dwarf AGNs given the at least $\sim0.3$ dex scatter in the \citet{Burke2021} relation and large uncertainties on the individual damping timescale measurements ($\sim0.4$ dex or larger depending on the ratio of the damping timescale to the light curve baseline), so results in much fewer candidates compared to the stellar mass criteria. Indeed, if we relax the damping timescale constraint to rest-frame $\log( \tau_{\rm{DRW}} {\rm{/ days}})<2.3$ with no floor at the light curve cadence, we find $\sim150$ sources. We found 17 sources with observed-frame damping timescales less than 7 days, which could be interpreted as an upper limit on the true damping timescale or sources whose variability is statistically spurious or not characteristic of an AGN.

\begin{table*}
\caption{Full Catalog of \Nvar{} variable AGN candidates regardless of stellar mass or variability timescale. All coordinates are given in the J2000 epoch. Values of $-1$ indicate invalid values. The first 5 rows are shown for formatting guidance. A full version of this table is available in the online version. \label{tab:catalogall}}
\tiny
\begin{tabular}{ccccccccccccccc}
\hline
Name & RA & dec & $g$ & SNR & $\sigma_{\rm{LB}}$ & $\log\ \frac{M_{\ast}}{M_{\odot}}$ & $\Delta\chi_\nu^2$ & $\log\ \frac{\tau_{\rm{\textsc{DRW},rest}}}{\rm{days}}$ & $z_{\rm{ph}}$ & $z_{\rm{sp}}$ & $z_{\rm{sp}}$ source & $W1{-}W2$ & $\log \frac{F_{2-7 {\rm{\ keV}}}}{{\rm{erg\  s}}^{-1} {\rm{\ cm}}^{-2}}$ & Field \\
 & (deg) & (deg) & (mag) & & & (dex) & & (dex) & & & & & (dex) &  \\
\hline
032833.79-271056.42 & 52.1408 & -27.1823 &     23.7 & 14.2 &      13.3 & $10.1 \pm 0.4$ &    0.1 &    $3.2 \pm 0.6$ & $1.02^{+0.03}_{-0.04}$ &     NaN &     None &    0.4 &    NaN & SN-C3 \\
J032951.20-271057.43 & 52.4633 & -27.1826 &     23.4 & 15.3 &      12.7 &  $7.8 \pm 0.7$ &    0.1 &    $1.9 \pm 0.3$ & $0.24^{+0.04}_{-0.02}$ &     NaN &     None &   18.6 &    NaN & SN-C3 \\
J032944.64-271107.53 & 52.4360 & -27.1854 &     23.4 & 20.1 &      16.3 &  $9.4 \pm 0.5$ &    0.0 &    $2.4 \pm 0.8$ & $1.09^{+0.02}_{-0.02}$ &     NaN &     None &    0.9 &    NaN & SN-C3 \\
J032845.45-271117.18 & 52.1894 & -27.1881 &     21.5 & 22.8 &      18.3 &  $9.7 \pm 0.0$ &    5.0 &    $1.7 \pm 0.2$ & $1.30^{+0.10}_{-0.15}$ &     NaN &     None &    1.1 &    NaN & SN-C3 \\
J032843.19-271117.75 & 52.1799 & -27.1883 &     23.8 & 13.4 &      12.4 &  $8.8 \pm 0.5$ &    0.0 &    $2.7 \pm 0.8$ & $0.60^{+0.25}_{-0.05}$ &     NaN &     None &    NaN &    NaN & SN-C3 \\
$\cdots$ & $\cdots$ & $\cdots$ & $\cdots$ & $\cdots$ & $\cdots$ & $\cdots$ & $\cdots$ & $\cdots$ & $\cdots$ & $\cdots$ & $\cdots$ & $\cdots$ & $\cdots$ & $\cdots$ \\
\hline
\end{tabular}
\end{table*}

\begin{table*}
\caption{Catalog of \Nfinal{} variable AGN candidates with rapid variability ($\log\ (\tau_{\rm{DRW}} \rm{/ days}) < 1.5$). We consider these sources the best dwarf AGN candidates. All coordinates are given in the J2000 epoch. Values of $-1$ indicate invalid values. \label{tab:catalogvar}}
\tiny
\begin{tabular}{ccccccccccccccc}
\hline
Name & RA & dec & $g$ & SNR & $\sigma_{\rm{LB}}$ & $\log\ \frac{M_{\ast}}{M_{\odot}}$ & $\Delta\chi_\nu^2$ & $\log\ \frac{\tau_{\rm{\textsc{DRW},rest}}}{\rm{days}}$ & $z_{\rm{ph}}$ & $z_{\rm{sp}}$ & $z_{\rm{sp}}$ source & $W1{-}W2$ & $\log \frac{F_{2-7 {\rm{\ keV}}}}{{\rm{erg\  s}}^{-1} {\rm{\ cm}}^{-2}}$ & Field \\
 & (deg) & (deg) & (mag) & & & (dex) & & (dex) & & & & & (dex) &  \\
\hline
J033150.63-282910.86 & 52.9610 & -28.4863 &     23.7 & 11.6 &      10.8 & $10.4 \pm 0.3$ &    0.4 & $1.2 \pm 0.2$ & $1.02^{+0.02}_{-0.03}$ &     NaN &     None &   22.5 & NaN & SN-C3 \\
J033129.06-272336.65 & 52.8711 & -27.3935 &     21.6 & 13.9 &      10.4 &  $9.4 \pm 0.3$ &    0.4 & $1.2 \pm 0.2$ & $0.36^{+0.02}_{-0.03}$ &  0.3456 &     2dF  &    0.6 & NaN & SN-C3 \\
J033051.65-272856.18 & 52.7152 & -27.4823 &     21.0 &  7.0 &       8.5 & $11.1 \pm 0.1$ &    5.5 & $1.3 \pm 0.2$ & $0.58^{+0.02}_{-0.02}$ &     NaN &     None &    0.5 & NaN & SN-C3 \\
J033002.95-273248.41 & 52.5123 & -27.5468 &     20.1 & 21.8 &      14.1 & $10.0 \pm 0.5$ &    0.8 & $1.4 \pm 0.2$ & $0.55^{+0.15}_{-0.20}$ &  0.5270 &     2dF  &    0.7 & NaN & SN-C3 \\
J032723.33-275657.10 & 51.8472 & -27.9492 &     21.1 & 20.8 &      16.4 & $10.5 \pm 0.2$ &    1.1 & $1.5 \pm 0.2$ & $0.35^{+0.01}_{-0.02}$ &  0.4635 &     2dF  &    0.6 & NaN & SN-C3 \\
J033208.67-273112.08 & 53.0361 & -27.5200 &     22.9 &  4.8 &       7.5 &  $9.8 \pm 0.4$ &    0.4 & $1.3 \pm 0.3$ & $1.30^{+0.10}_{-0.15}$ &     NaN &     None &   17.8 & NaN & SN-C3 \\
J033226.49-280520.08 & 53.1104 & -28.0889 &     23.3 & 12.2 &       4.9 &  $6.4 \pm 0.6$ &    0.0 & $1.5 \pm 0.7$ & $0.04^{+0.01}_{-0.03}$ &     NaN &     None &    NaN & NaN & SN-C3 \\
J033240.53-283622.28 & 53.1689 & -28.6062 &     20.9 &  9.7 &       5.1 &  $9.9 \pm 0.2$ &    1.1 & $1.3 \pm 0.4$ & $0.31^{+0.04}_{-0.05}$ &  0.3224 &     2dF  &    0.3 & NaN & SN-C3 \\
J032955.16-284359.67 & 52.4798 & -28.7332 &     20.1 &  8.5 &       5.4 &  $8.6 \pm 0.1$ &   10.7 & $1.5 \pm 0.4$ & $0.04^{+0.02}_{-0.03}$ &     NaN &     None &   -0.1 & NaN & SN-C3 \\
J032705.57-284011.91 & 51.7732 & -28.6700 &     21.7 & 16.3 &      10.4 & $10.7 \pm 0.3$ &    0.2 & $1.4 \pm 0.2$ & $0.71^{+0.01}_{-0.07}$ &     NaN &     None &    0.2 & NaN & SN-C3 \\
J003526.35-443806.37 &  8.8598 & -44.6351 &     22.4 & 10.2 &       7.4 &  $9.4 \pm 0.3$ &    0.7 & $1.4 \pm 0.2$ & $1.15^{+0.15}_{-0.50}$ &     NaN &     None &   17.3 & NaN & SN-E2 \\
J004020.38-432053.86 & 10.0849 & -43.3483 &     19.2 & 13.2 &       6.1 & $10.3 \pm 0.1$ &    3.5 & $0.9 \pm 0.1$ & $0.90^{+0.20}_{-0.25}$ &     NaN &     None &    1.2 & NaN & SN-E2 \\
J003705.78-444006.02 &  9.2741 & -44.6683 &     21.4 &  8.3 &       4.7 &  $8.0 \pm 0.5$ &    0.4 & $1.2 \pm 0.2$ & $0.77^{+0.04}_{-0.02}$ &     NaN &     None &    0.2 & NaN & SN-E2 \\
J022305.26-042800.90 & 35.7719 &  -4.4669 &     22.5 & 10.6 &       9.2 &  $9.1 \pm 0.3$ &    0.6 & $1.4 \pm 0.2$ & $0.90^{+0.25}_{-0.10}$ &  0.8194 &     SDSS &    0.7 & NaN & SN-X3 \\
J022446.71-050017.47 & 36.1946 &  -5.0049 &     19.7 &  4.0 &       4.8 &  $7.7 \pm 0.2$ &    3.1 & $1.3 \pm 0.3$ & $0.05^{+0.05}_{-0.04}$ &  0.0694 &     PanS &    0.1 & NaN & SN-X3 \\
\hline
\end{tabular}
\end{table*}

\subsection{Spectroscopic Properties}

Three of our sources with rapid variability in Table~\ref{tab:catalogvar}, J033129.06$-$272336.65, J032723.33$-$275657.10, and J033240.53$-$283622.28 have 2dF spectra from the OzDES\footnote{Australian Dark Energy Survey} program \citep{Lidman2020}. One source, J033129.06-272336.65, shows possible absorption features but no clear emission lines, which may indicate the AGN emission is diluted by absorption from an old stellar population. J032723.33$-$275657.10 and J033240.53$-$283622.28 do not show strong emission or absorption features. These sources are faint $g \sim 21$, and probably require higher S/N spectra. Another source, J022446.71$-$050017.47, has a spectrum from the Pan-STARRS SN sample \citep{Rest2014}, but its spectrum would be contaminated by SN emission.

\subsubsection{J022305.3-042800.9}

\begin{figure*}
\includegraphics[width=0.98\textwidth]{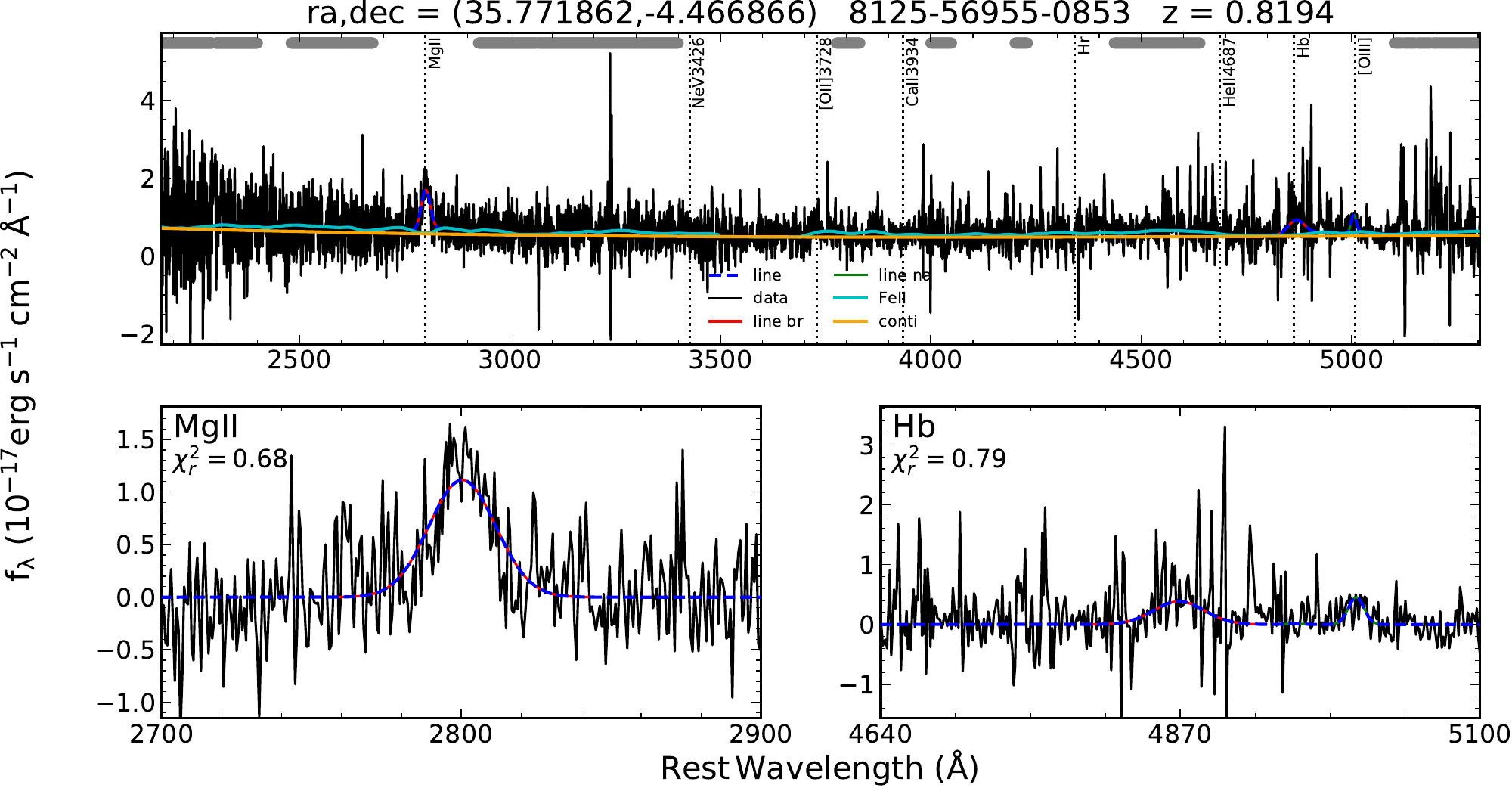}
\centering
\caption{Optical spectrum for example source J022305.3$-$042800.9 from SDSS. A global fitting is applied to the spectrum having subtracted the host component in the upper panel. A power-law plus 3rd-order polynomial and Gaussians are used to fit the continuum and emission lines, respectively. The grey bands on the top are line-free windows selected to determine the continuum emission. The lower panels show the zoomed-in emission line regions of \MgII\ and \hbeta . Broad \MgII\ and broad \hbeta\ are both detected at the 3.5$\sigma$ and 1.2$\sigma$ significance levels, respectively, yielding virial BH mass of $\log(M_{\rm{BH}}/M_{\odot}) = 6.4\pm0.6\ M_{\odot}$ using \hbeta\ and $\log(M_{\rm{BH}}/M_{\odot}) = 6.6\pm0.1$ using \MgII .
\label{fig:specex}}
\end{figure*}

We identified one source with short-timescale variability in Table~\ref{tab:catalogvar}, J022305.3-042800.9 (see Figure~\ref{fig:examples}), with a good SDSS spectrum. The stellar mass of the source is $M_{\ast} = 10^{9.1\pm0.3}\ M_{\odot}$ and its $z_{\rm{ph}}=0.90^{+0.25}_{-0.10}$ is consistent with the spectroscopic redshift of $z_{\rm{sp}}=0.8194$. The calibrated SDSS spectra enables straightforward spectral modeling using existing tools. This source is of class ``QSO'' and shows broad H$\beta$ and Mg~\textsc{ii} emission lines. We will use the broad lines to obtain single-epoch virial BH mass estimates.

To determine the significance of the broad emission lines and to measure their profiles for virial BH mass estimates, we fit spectral models following the procedures as described in detail in \citet{Shen2019} using the software \textsc{PyQSOFit}\footnote{\url{https://github.com/legolason/PyQSOFit}} \citep{Guo2018}. The model is a linear combination of a power-law continuum, a 3rd-order polynomial (to account for reddening), a pseudo continuum constructed from Fe\,II emission templates, and single or multiple Gaussians for the emission lines. Since uncertainties in the continuum model may induce subtle effects on measurements for weak emission lines, we first perform a global fit to the emission-line free region to better quantify the continuum. We then fit multiple Gaussian models to the continuum-subtracted spectrum around the broad emission line region locally. 

More specifically, we model the \MgII\ and \hbeta\ lines each using a single broad (FWHM $ > 1200$ km s$^{-1}$) Gaussian component. Given the low S/N of the spectrum, adding additional components does not improve the fit significantly. The \OIII\ $\lambda$5007 \AA\ emission line appears affected by sky line residuals. Therefore, we are unable to use it as a template to tie the narrow components. This contributes to the large uncertainties in our virial mass measurements. We use 100 Monte Carlo simulations to estimate the uncertainty in the line measurements.

Our spectral modeling is shown in Figure~\ref{fig:specex}. Using the virial mass relation \citep{Shen2013}, 
\begin{equation}
    \log\left(\frac{M_{\rm{BH}}}{M_{\odot}}\right) = a + b \log\left(\frac{\lambda L_{\lambda}}{10^{44} {\rm{\ erg\ s}}^{-1} }\right) + 2 \log\left(\frac{\rm{FWHM}}{\rm{km\ s}^{-1}}\right) ,
\end{equation}
we obtain BH mass estimates of $\log(M_{\rm{BH}}/M_{\odot}) = 6.4\pm0.6$ (H$\beta$) and $\log(M_{\rm{BH}}/M_{\odot}) = 6.6\pm0.1$ \MgII\ using the relations of \citet{Mejia-Restrepo2016} with $(a, b) = (0.864, 0.568)$ (\hbeta ) and $(a, b) = (0.955, 0.599)$ (\MgII ). The redshift and BH mass are similar to the source of \citet{Guo2018}, which is not within one of the deep NIR fields of this work. Although the BH mass is somewhat larger than other samples (e.g., \citealt{Reines2013}), those samples are generally limited to low redshifts ($z\lesssim0.15$). This demonstrates the capability of variability to identify lower-mass SMBHs at intermediate redshifts. Both values are consistent with the predicted $\log(M_{\rm{BH}}/M_{\odot})  = 6.4 \pm 0.4$ from the $M_{\rm{BH}} - \tau_{\rm{DRW}}$ of \citet{Burke2021}.

\section{Discussion} \label{sec:discussion}

\subsection{Comparison to Previous Works}

\begin{figure}
\centering
{\includegraphics[width=0.5\textwidth]{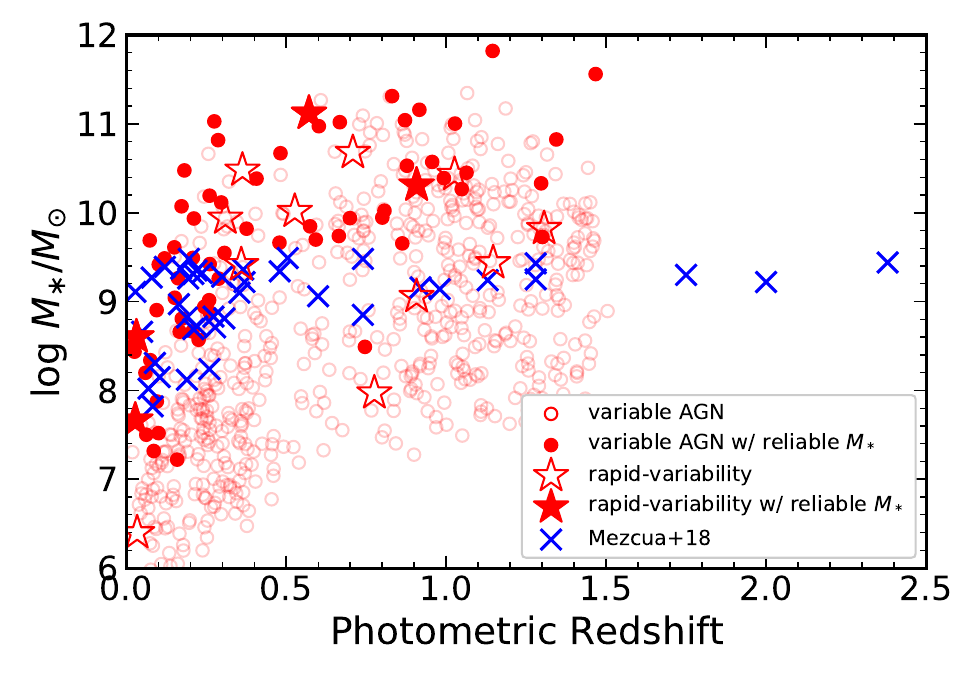}
}
\caption{Stellar mass versus photometric redshift for variable galaxies in our DES deep field sample with reliable stellar mass estimate (red). The sources with rapid optical variability in Table~\ref{tab:catalogvar} are shown as red star symbols. The typical (mean) uncertainties in stellar mass and redshift are shown at the bottom of the plot. For comparison, we show the X-ray-selected dwarf AGNs from \emph{Chandra} COSMOS legacy survey imaging \citep{Mezcua2018}. \label{fig:redshiftmezcua}}
\end{figure}

\cite{Baldassare2018} and \citet{Baldassare2020} used the \cite{Butler2011} method to select dwarf AGNs $(M_{\ast}<10^{10} M_{\odot})$ in SDSS Stripe 82 and Palomar Transient Factory light curves respectively using difference imaging. The variable AGN fraction found in these works is broadly consistent with our findings ($\lesssim1\%$) given the different cadence, baselines, and photometric precision between surveys. A similar work was performed using Zwicky Transient Facility imaging \citep{Ward2021b}. We have extended the findings of \citet{Baldassare2020}, showing that the occupation fraction of variable AGNs may be constant down to $M_{\ast}\sim10^7 M_{\odot}$. These earlier studies are restricted to samples of galaxies at $z<0.15$ with secure spectroscopic redshifts. In this work, we are able to extend the results to a sample at higher redshifts than \citet{Baldassare2018,Baldassare2020}. Although, we caution that our stellar masses are likely more uncertain due to the larger uncertainties in the photometric redshifts and lack of deep UV and MIR photometry.

As an additional point of comparison, we plot the stellar mass versus redshift against the sample of dwarf AGN selected from the \emph{Chandra} COSMOS legacy survey \citep{Mezcua2018} in Figure~\ref{fig:redshiftmezcua}. This demonstrates that our variability-selected AGN have comparable redshifts and stellar masses to the deep X-ray AGN selection technique. This is consistent with Figure~7 of \citet{Guo2020}, which showed that DES-SN variability selection can only be matched by deep X-ray/radio imaging in the mass-redshift parameter space. In contrast, our technique covers a much larger area of the sky, but will miss Type II AGN with variability obscured in the optical. 

\subsection{X-ray Properties}

\begin{figure}
{\includegraphics[width=0.5\textwidth]{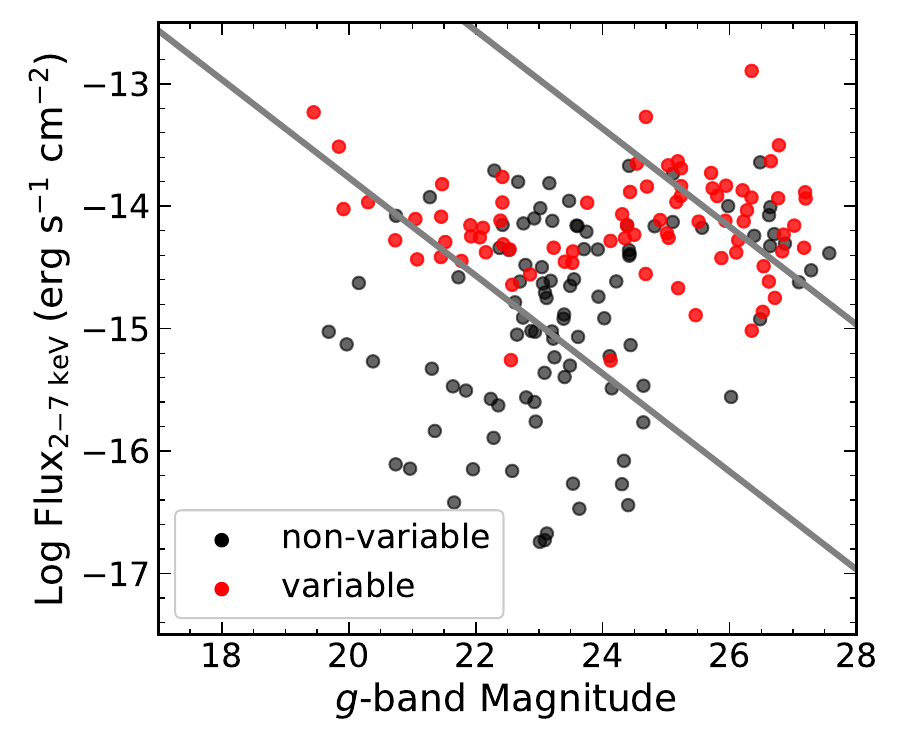}
}
\caption{\emph{Chandra} hard ($2-7$ keV) X-ray flux versus \emph{g}-band magnitude for variable (red) and non-variable (black) galaxies. The gray lines correspond to X-ray to optical flux ratios of $X/O=\pm1$. \label{fig:csc}}
\end{figure}

\begin{figure}
{\includegraphics[width=0.5\textwidth]{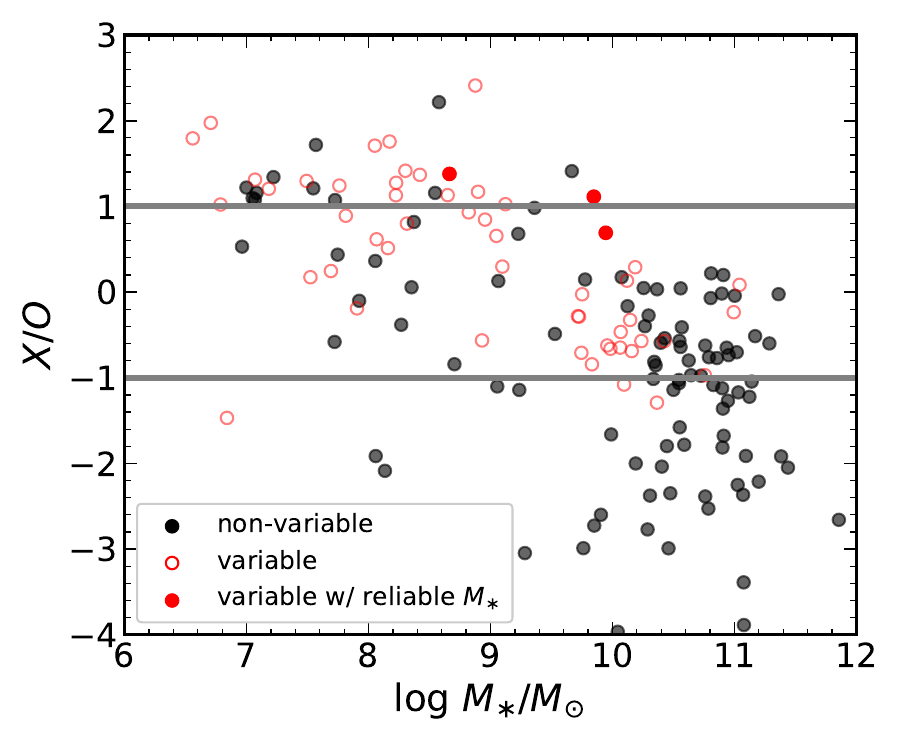}
}
\caption{X-ray to optical flux ratio versus stellar mass for variable (red) and non-variable (black) galaxies. The gray lines correspond to X-ray to optical flux ratios of $X/O=\pm1$. \label{fig:cscmass}}
\end{figure}

We match our sources to the \emph{Chandra} Source Catalog 2.0 \citep{Evans2010} of X-ray sources. This catalog includes detections in stacked observations in the \emph{Chandra} Deep Field South with a 5.8 Ms total exposure time. We matched our DES sources to this catalog using a 0.5$^{\prime\prime}$ radius. We found \NXray{} X-ray sources with variable DES light curves. 

We compute the X-ray to optical flux ratio $X/O$ of our sources in the \emph{Chandra} Deep Field South. We use the definition of \citet{Maccacaro1988}: 
\begin{equation}
    X/O = \log{(f_X/f_{\rm{opt}})} = \log f_x + \frac{\rm{mag_{opt}}}{2.5} + C
\end{equation}
where $f_{\rm{X}}$ is the X-ray flux (we use the \emph{Chandra} hard X-ray band $2{-}7$ keV), $\rm{mag_{opt}}$ is the optical magnitude (we use the \emph{g}-band deep coadd photometry), and $C$ is a zero-point constant ($C=4.77$ for the $g$ band). The results are shown in Figure~\ref{fig:csc} and Figure~\ref{fig:cscmass} for sources with acceptable stellar mass estimates. The non-variable sources with X-ray detections are likely optically-obscured AGN. In addition, sources may have significant contamination from star-formation. Nevertheless, the fact that all variable sources fall near or above the $X/O=-1$ line reassures that the variability in most of our sources is of AGN nature.

Only three of the AGN candidates with rapid optical variability timescales in Table~\ref{tab:catalogvar} appear to be on an archival \emph{Chandra} observation, and none of those candidates are present in the \emph{Chandra} Source Catalog 2.0. As such we search for X-ray counterparts in the observations of another X-ray telescope, {\em XMM}-Newton, by using the processed data and region files of the {\em XMM} cluster survey \citep[XCS,][]{xcsfoundation}. XCS is a serendipitous survey of the {\em XMM} data archive that is primarily focused on the measurement of galaxy cluster properties, but also locates and catalogues X-ray point sources. The XCS source finder (XAPA) first locates X-ray sources in {\em XMM} data, then classifies their emission as point or extended.

We use X-ray: Generate and Analyse \citep[\texttt{XGA}\footnote{\href{https://github.com/DavidT3/XGA}{X-ray: Generate and Analyse GitHub}},][]{xgapaper}; a new, open-source, X-ray astronomy analysis module developed by XCS, to first determine which of the candidates in Table~\ref{tab:catalogvar} have {\em XMM} data, and then which of those candidates match to an XCS point source. We find that all of the 11 candidates appear in at least one {\em XMM} observation, with the most well observed candidates appearing in eight. We also find that 6 of the 11 have a corresponding XCS point source match in at least one {\em XMM} observation, where we define a match as the DES coordinate falling within an XCS point source region. We use \texttt{XGA} to generate stacked {\em XMM} count-rate maps, both with and without spatially varying PSF correction; as the sample in Table~\ref{tab:catalogvar} is small, visual inspection to confirm the veracity of the matches is possible. In one case we note what appears to be very faint point source emission at the coordinates of the candidate, with no corresponding XCS source region. In all other cases the detections and non-detections appear to be appropriate. The notebook containing the brief {\em XMM} analysis is available on GitHub\footnote{\href{https://github.com/DavidT3/XCS-DES-AGN-Prelim}{{\em XMM} Exploration Jupyter Notebook}}.

We also check the 4XMM DR11 catalogue \citep[][]{fourxmmdr11}, and find that 7 of the 11 candidates match to a 4XMM source within 3\arcsec, 6 of which are detected by XCS. The additional candidate detected by 4XMM is the same one for which we note a slight emission during visual inspection. The lack of detection is likely caused by XCS performing source finding in the 0.5-2.0~keV band, which is optimised for the detection of galaxy clusters. We defer the full X-ray analysis of our variability-selected AGN sample to a future paper.

\subsection{\emph{WISE} Properties}

\begin{figure}
{\includegraphics[width=0.5\textwidth]{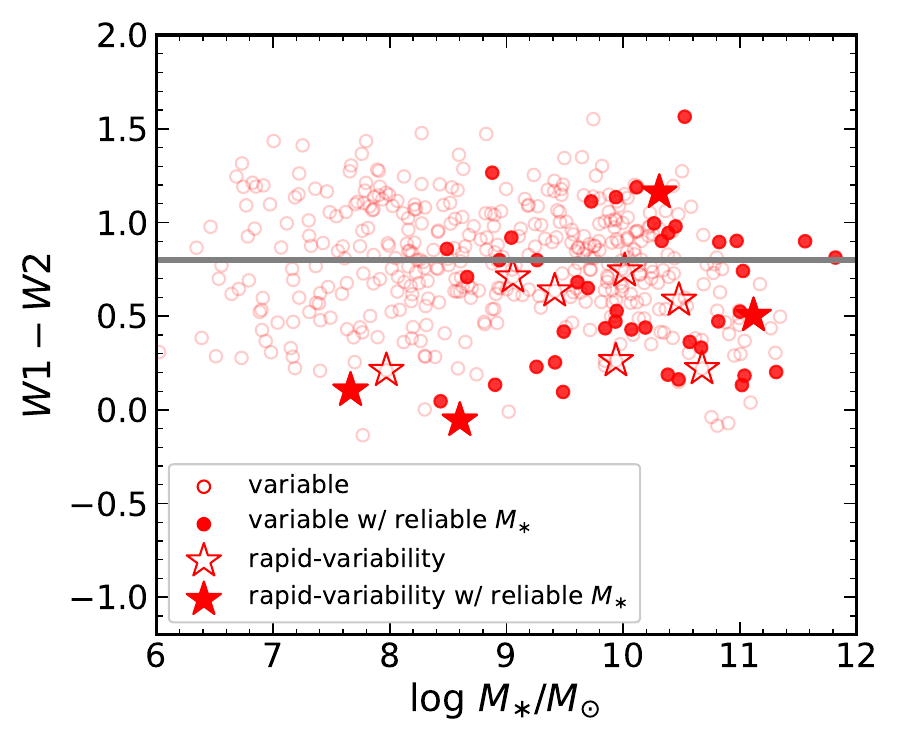}
}
\caption{\emph{WISE} $W1-W2$ colors versus stellar mass for variable sources. The grey line is the simple threshold for AGN selection of $W1-W2>0.8$ proposed by \citet{Stern2012}. \label{fig:wise}}
\end{figure}

We match our variable AGN candidates to the all-sky unWISE extra-galactic catalog \citep{Schlafly2019} using a 5$^{\prime\prime}$ radius which includes both galaxies and AGNs. We find \NWISEAGN{} matches that satisfy the simple \emph{WISE} AGN selection criteria of $W1-W2>0.8$ \citep{Stern2012} out of \NWISE{} total matches. We plot the $W1-W2$ colors versus stellar mass in Figure~\ref{fig:wise} for sources with acceptable stellar mass estimates. The colors follow the upper-tail of the $W1-W2$ distribution for galaxies and AGNs \citep{Stern2012,Assef2013}, suggesting some AGN contribution to the mid-infrared (MIR) emission. Nevertheless, this demonstrates that a large fraction ($\sim50$ percent) of our variable AGN sources are not dominated by AGN emission in the MIR. Hence, these variable AGN would be missed by MIR color selection.

\subsection{Comparison to Host Scaling Relations}

\begin{figure}
\includegraphics[width=0.48\textwidth]{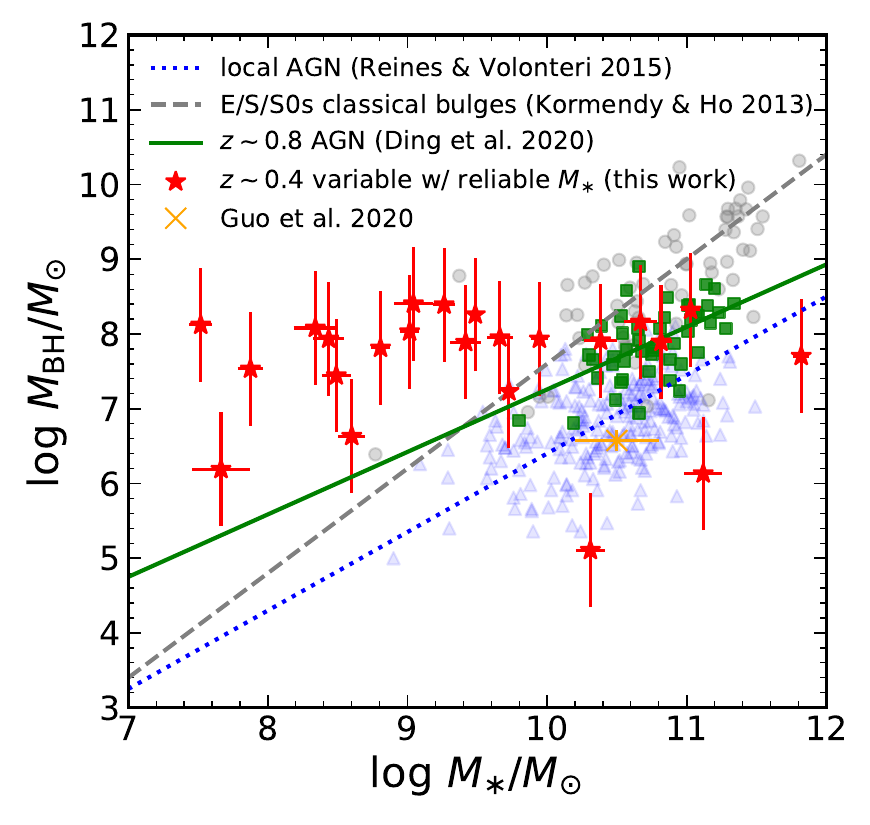}
\centering
\caption{Black hole masses estimated from the damping timescales using the relation of \citet{Burke2021} versus host galaxy stellar mass $M_{\ast}$ estimates from SED fitting for our variable sources with constrained variability timescales and reliable stellar mass estimates (red star symbols). For comparison, we show the scaling relations from X-ray selected intermediate-redshift AGNs and local samples of AGNs and inactive galaxies. The green solid line shows the best-fit relation of the sample of 48 X-ray selected AGNs with a median $z\sim0.8$ from \citet{Cisternas2011} and \citet{Schramm2013} re-analyzed by \citet{Ding2020}, with individual sources shown as green square symbols. The blue dotted line represents the best-fit relation in local AGNs from \citet{Reines2015}, with individual sources shown as blue triangle symbols. The gray dashed line denotes the best-fit relation using the sample of ellipticals and spiral/S0 galaxies with classical bulges from \citet{KormendyHo2013}, with individual sources shown as gray circle symbols. The DES variability-selected dwarf AGN from \citet{Guo2020} is shown as the orange `$\times$' symbol. The error bars on the red points are statistical uncertainties.
\label{fig:massrelation}}
\end{figure}

We show the black hole masses estimated from the damping timescales using the relation of \citet{Burke2021} versus host galaxy stellar mass estimates from \textsc{cigale} SED fitting for our \Nfinal{} sources with rapid optical variability in Table~\ref{tab:catalogvar} in Figure~\ref{fig:massrelation}. Shown for comparison is best-fitting relation from the X-ray selected AGN sample at median $z\sim0.8$ from \citet{Cisternas2011,Schramm2013} re-analyzed by \citet{Ding2020}. The virial black hole masses were estimated based on single-epoch spectra using broad H$\beta$ and/or broad \MgII . The comparison sample includes 32 objects from \citet{Cisternas2011} and 16 objects from \citet{Schramm2013}. The total stellar masses of the \citet{Cisternas2011} sample were estimated by the empirical relation between $M_{*}/L$ and redshift and luminosity in the Hubble Space Telescope (HST) F814W band, which was established using a sample of 199 AGN host galaxies. The total stellar masses for the \citet{Schramm2013} sample were estimated from the galaxy absolute magnitude $M_{V}$ and rest-frame $(B-V)$ color measured from HST imaging for quasar-host decomposition using the $M/L$ calibration of \citet{Bell2003}. 

Also shown for context in Figure~\ref{fig:massrelation} are the best-fit scaling relations for local samples of inactive galaxies \citep[e.g.,][]{Haring2004,KormendyHo2013,McConnell2013} and low-redshift AGNs \citep{Reines2015}. Our sample appears broadly consistent with the relation of \citet{Reines2015} or the $z\sim0.8$ AGNs. This is not unexpected given the typical redshfit of our sources of $z\sim0.4$, in-between the intermediate-redshift and local AGN populations. However, any apparent offset is likely insignificant accounting for possible systematic uncertainties in the stellar mass estimates. Also, by imposing the cut on the variability timescale, we introduce a selection effect which may reduce the correlation. While based on only 11 data points and the results of \citet{Guo2020}, our results may suggest no significant redshift evolution in the $M_{\rm{BH}}$--$M_{\ast}$ scaling relation from redshift $z\sim1$ to $z\sim0$ \citep[see also][]{Ding2020,Li2021}, which is consistent with previous results based on the $M_{\rm{BH}}$--$\sigma_{\ast}$ relation \citep[e.g.,][]{Shen2015,Sexton2019}. 

\section{Conclusions} \label{sec:conclusions}

We have identified \Nvar{} candidate AGNs at $z<1.5$ in the DES deep fields using optical variability. Using SED fitting for stellar mass estimation, we found \Ndwarf{} candidate dwarf AGNs with host stellar mass $M_{\ast} < 10^{9.5}\ M_{\odot}$ (at a median photometric redshift of $\langle z \rangle \sim 0.9$) and \Nfinal{} candidates with short-timescale variability $\langle z \rangle \sim 0.4$. Our dwarf AGNs are at higher redshift at a given stellar mass than previous variability-selected dwarf AGN samples, and on-par with dwarf AGN identification in deep X-ray/radio surveys. Such dwarf AGNs at these intermediate redshifts are more likely to be pristine analogs of SMBH seeds that formed at high redshift. We measure the variable AGN fraction in our parent galaxy sample of \Nparent{} objects, which, consistent with previous work, depends on stellar mass due to a variety of selection effects \citep{Burke2022}. However, we caution that our sample is likely to contain some false positives given the AGN/star-formation degeneracies in stellar mass estimates and scatter in the black hole -- host galaxy stellar mass relation. Our candidates require further follow-up to measure their black hole masses.

Analysis of the X-ray and MIR fluxes in most of our variable sources is consistent with their AGN nature. However, their host stellar mass estimates remain somewhat uncertain given the limitations of optical and near IR SED fitting and contamination from the AGN emission. Nevertheless, our catalog of variable intermediate-redshift dwarf AGNs with high-quality optical light curves in legacy fields probes a unique parameter space of dwarf AGN searches. Extension to deep field public catalogs with uniformly-extracted photometry from the UV to MIR (e.g., \citealt{Davies2021}) will help constrain the stellar mass estimates. Future deep imaging surveys in the IR and UV will also help constrain the stellar masses at higher redshifts. High photometric precision and higher cadence light curves will enable detection and mass estimation of IMBHs using the relation of \citet{Burke2021}. Continued monitoring in these deep fields is ongoing with DECam to further extend the light curve duration, which will enable more robust measurements of the long-term AGN optical variability damping timescale \citep{Kozlowski2021} for more accurate black hole mass estimation. Surveys with a more rapid cadence will enable smaller variability timescale measurements and probe even lower black hole masses \citep{Bellm2021}. These data will be essential for enabling an accurate determination of the AGN occupation fraction in low-mass galaxies in the era of the Rubin Observatory \citep{Ivezic2019}.

\section*{Acknowledgements}

C.J.B. and Y.-C.C. acknowledge support from the Illinois Graduate Survey Science Fellowship. X.L., Y.S., and Y.-C.C. acknowledge support by NSF grant AST-2108162 and AST-2206499. Y.S. acknowledges support by NSF grant AST-2009947. This research was supported in part by the National Science Foundation under PHY-1748958.

We thank Brian Yanny and Ken Herner for help with the DES computing resources used for this work. We thank Richard Kessler for help obtaining the original DES-SN program light curves and for comments which improved the clarity of the text. 

Funding for the DES Projects has been provided by the U.S. Department of Energy, the U.S. National Science Foundation, the Ministry of Science and Education of Spain, 
the Science and Technology Facilities Council of the United Kingdom, the Higher Education Funding Council for England, the National Center for Supercomputing 
Applications at the University of Illinois at Urbana-Champaign, the Kavli Institute of Cosmological Physics at the University of Chicago, 
the Center for Cosmology and Astro-Particle Physics at the Ohio State University,
the Mitchell Institute for Fundamental Physics and Astronomy at Texas A\&M University, Financiadora de Estudos e Projetos, 
Funda{\c c}{\~a}o Carlos Chagas Filho de Amparo {\`a} Pesquisa do Estado do Rio de Janeiro, Conselho Nacional de Desenvolvimento Cient{\'i}fico e Tecnol{\'o}gico and 
the Minist{\'e}rio da Ci{\^e}ncia, Tecnologia e Inova{\c c}{\~a}o, the Deutsche Forschungsgemeinschaft and the Collaborating Institutions in the Dark Energy Survey. 

The Collaborating Institutions are Argonne National Laboratory, the University of California at Santa Cruz, the University of Cambridge, Centro de Investigaciones Energ{\'e}ticas, 
Medioambientales y Tecnol{\'o}gicas-Madrid, the University of Chicago, University College London, the DES-Brazil Consortium, the University of Edinburgh, 
the Eidgen{\"o}ssische Technische Hochschule (ETH) Z{\"u}rich, 
Fermi National Accelerator Laboratory, the University of Illinois at Urbana-Champaign, the Institut de Ci{\`e}ncies de l'Espai (IEEC/CSIC), 
the Institut de F{\'i}sica d'Altes Energies, Lawrence Berkeley National Laboratory, the Ludwig-Maximilians Universit{\"a}t M{\"u}nchen and the associated Excellence Cluster Universe, 
the University of Michigan, the National Optical Astronomy Observatory, the University of Nottingham, The Ohio State University, the University of Pennsylvania, the University of Portsmouth, 
SLAC National Accelerator Laboratory, Stanford University, the University of Sussex, Texas A\&M University, and the OzDES Membership Consortium.

Based in part on observations at Cerro Tololo Inter-American Observatory, National Optical Astronomy Observatory, which is operated by the Association of 
Universities for Research in Astronomy (AURA) under a cooperative agreement with the National Science Foundation.

The DES data management system is supported by the National Science Foundation under Grant Numbers AST-1138766 and AST-1536171.
The DES participants from Spanish institutions are partially supported by MINECO under grants AYA2015-71825, ESP2015-66861, FPA2015-68048, SEV-2016-0588, SEV-2016-0597, and MDM-2015-0509, 
some of which include ERDF funds from the European Union. IFAE is partially funded by the CERCA program of the Generalitat de Catalunya.
Research leading to these results has received funding from the European Research
Council under the European Union's Seventh Framework Program (FP7/2007-2013) including ERC grant agreements 240672, 291329, and 306478.
We  acknowledge support from the Brazilian Instituto Nacional de Ci\^encia
e Tecnologia (INCT) e-Universe (CNPq grant 465376/2014-2).

This manuscript has been authored by Fermi Research Alliance, LLC under Contract No. DE-AC02-07CH11359 with the U.S. Department of Energy, Office of Science, Office of High Energy Physics.

\section*{Data Availability}

The deep field photomtery catalogue will be made available as part of the cosmology data products release, following the completion of the DES 3-year weak lensing and galaxy clustering cosmology work. When available, optical spectra can be found following the references in \citet{Hartley2020}. Supplementary catalogues (e.g., \emph{WISE}, \emph{Chandra}, ZFOURGE) are available following the references provided in the text. 



\bibliographystyle{mnras}
\bibliography{ref} 




\appendix

\section{Photometric Error Correction}
\label{sec:errorbarcorrection}

\begin{figure*}
\centering
\includegraphics[width=1.0\textwidth]{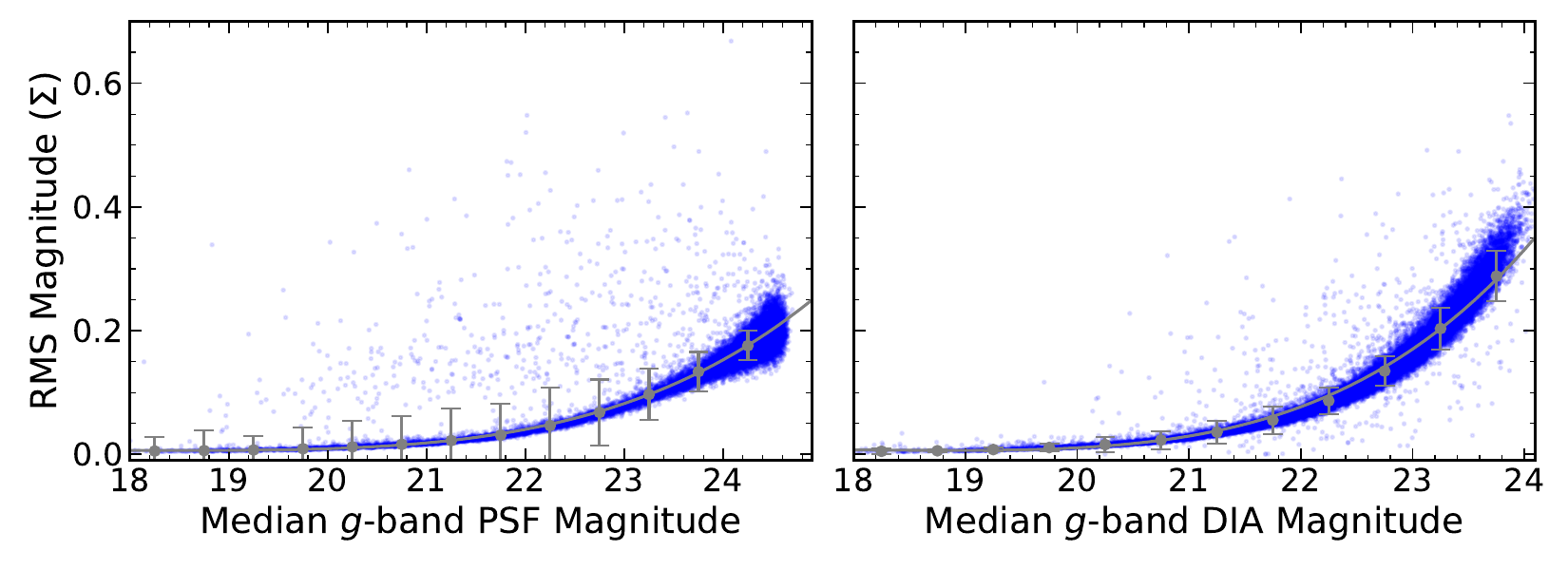}
\caption{RMS magnitude $\Sigma$ versus median \emph{g} band aperture magnitude for PSF (a) and DIA (b) SN-C3 light curves. The binned median and polynomial fit are shown in gray. The error bars are computed from the RMS of $\Sigma$ in each bin. \label{fig:correction}}
\end{figure*}

The uncertainties on the photometry derived from \textsc{SExtractor} do not include systematic sources of error, particularly relevant for fainter sources. To correct for systematic scatter in the photometry, we follow the method of \citet{Sesar2007}. We plot the RMS scatter of each light curve $\Sigma(m)$ as a function of median magnitude for each light curve using all sources in each field. Therefore, it is a good assumption that most sources are not intrinsically variable. We then compute the median of $\Sigma(m)$ in bins of width 0.5 magnitudes and fit a fourth-order polynomial through the binned medians with errors given by RMS of $\Sigma(m)$ in each bin. Assuming most sources are not intrinsically variable, the corrected errors on each measurement are given by:
\begin{equation}
    \sigma_i^\prime = \sqrt{\sigma_i^2 + \xi(m)^2}
\end{equation}
where $\sigma_i$ is the uncorrected error measurement and $\xi(m)$ is the fitted fourth order polynomial evaluated at magnitude $m$. We perform this correction separately for both PSF and DIA light curves. In general, DES has exceptionally stable photometry for sources brighter than $g\sim20$ but the RMS scatter increases as expected for fainter sources. We show the scatter versus magnitude in the SN-C3 field in Figure~\ref{fig:correction}. The scatter is larger in DIA light curves because of the various artifacts and systematic sources of noise introduced with difference imaging.

\section{Difference Imaging Zero Point Solution}
\label{sec:zeropoint}

\begin{figure}
\centering
\includegraphics[width=0.48\textwidth]{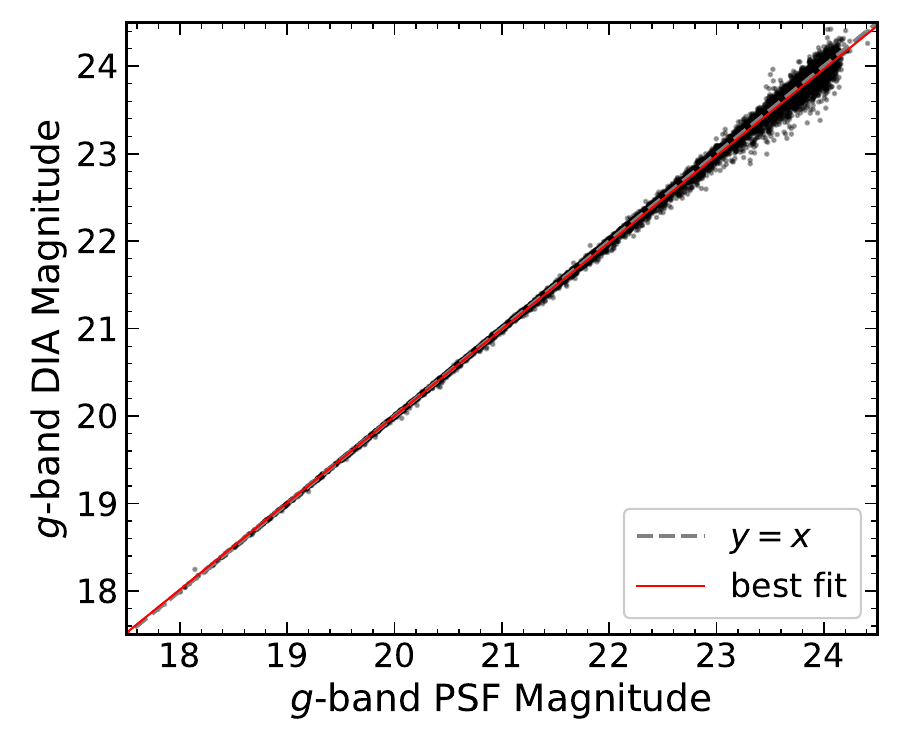}
\caption{PSF versus DIA magnitudes for unresolved SN-C3 sources after correcting for the zero point difference following the procedure described in Appendix~\ref{sec:zeropoint}. The resulting linear regression and line of $y=x$ are shown after correction for comparison. \label{fig:zeropoint}}
\end{figure}

To determine the zero point of our DIA magnitudes and to place our DIA photometry on the DES photometric system, we note that the DIA magnitudes with a $5^{\prime\prime}$ aperture should be equivalent to PSF magnitudes for unresolved sources. Therefore, in each field, we simply plot the median PSF versus DIA magnitudes for each unresolved source and perform linear regression. We take difference between the $y=x$ line and the fitted line and at $g=20$ as the DIA zero point solution. We show the result for SN-C3 sources in Figure~\ref{fig:zeropoint} after performing the zero point correction. The DIA magnitudes are tightly correlated with the PSF magnitudes, providing further validation of our difference imaging pipeline.

\section{Stellar Mass Estimate Validation}
\label{sec:stellarmassvalidation}

\begin{figure*}
\centering
\includegraphics[width=1\textwidth]{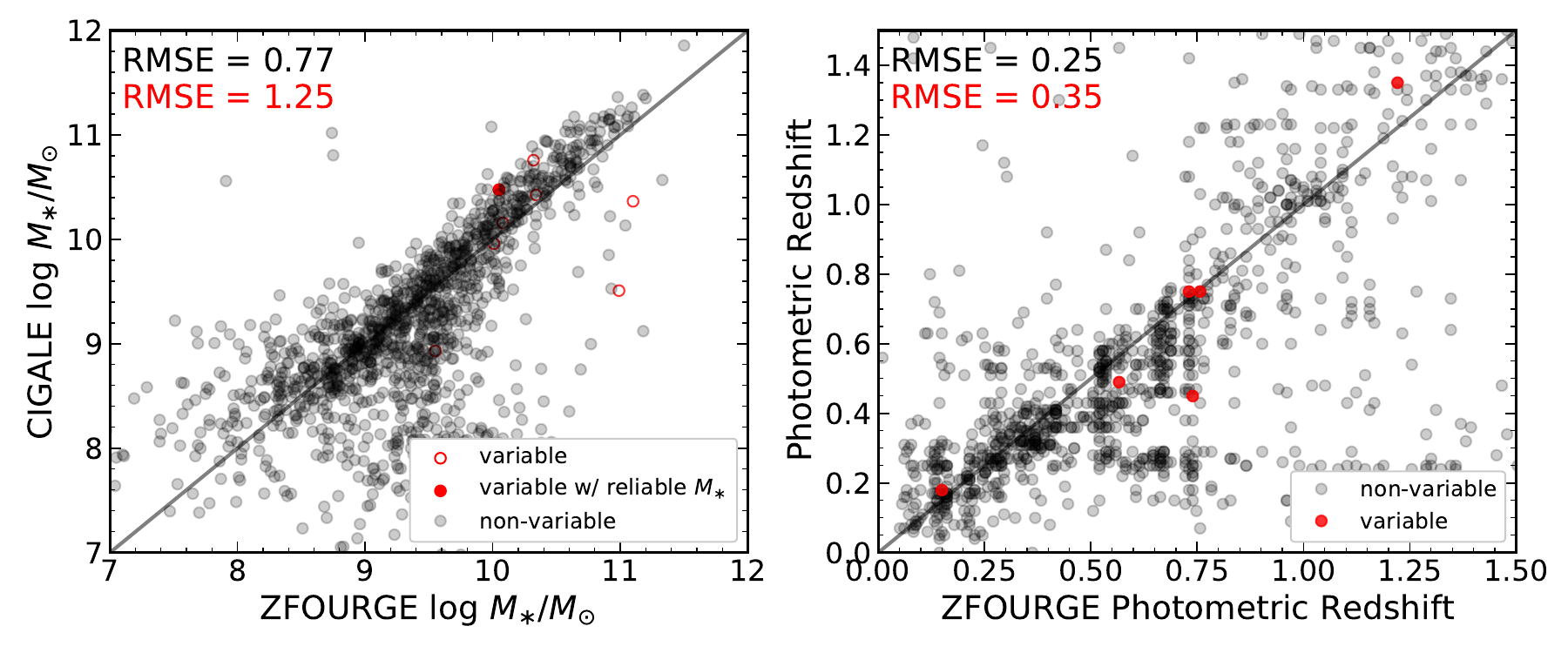}
\caption{Stellar mass and photometric redshift comparison between our results and ZFOURGE. The root mean square error (RMSE) for the variable (red) and non-variable (black) sources are shown in the upper-left hand corner of each panel. \label{fig:zfourge}}
\end{figure*}

To validate our \textsc{cigale} stellar mass estimates described in \S\ref{sec:stellarmass}, we compare our \textsc{cigale} stellar masses to ZFOURGE \citep{Tomczak2014,Straatman2016}. ZFOURGE is a deep medium-band imaging survey which provides an observational benchmark of galaxy properties at intermediate redshift. After matching our SN-C3 sources to ZFOURGE-CDFS sources, we compare the ZFOURGE results to our stellar masses from \textsc{cigale} and photometric redshifts from the \citet{Yang2017} method in Figure~\ref{fig:zfourge}. We find that our \textsc{cigale} stellar mass and redshifts are well-correlated with ZFOURGE. We find a root mean square error (RMSE) of \RMSEmassnonvar{} dex for non-variable galaxies. For the photometric redshift comparison, we compute RMSE($\Delta z$), where $\Delta z = |z_{\rm{ph}} - z_{\rm{ZFOURGE}}|/(1 + z_{\rm{ZFOURGE}})$. We note that ZFOURGE does not consider an AGN component in their SED model, so the photometric redshift and stellar mass comparison of galaxies with luminous AGNs should be treated with caution.

This RMSE is consistent with previous findings that stellar mass estimates can include systematic scatter of up to 20 percent due to differences in model assumptions, even when the photometric redshift is accurate \citep{Ciesla2015,Boquien2019}. In particular, we caution that star-formation is often degenerate with UV/optical AGN emission from the accretion disk. We attempt to break this degeneracy by using variability information as a simple prior on the inclination angle of the standard SKIRTOR model AGN in \textsc{cigale} (variable/Type-I: $i<30$ deg; non-variable/Type-II: $i>30$ deg) but allow the AGN luminosity fraction to vary between 0.1 and 0.9 in either case. 

\begin{figure*}
\centering
\includegraphics[width=1\textwidth]{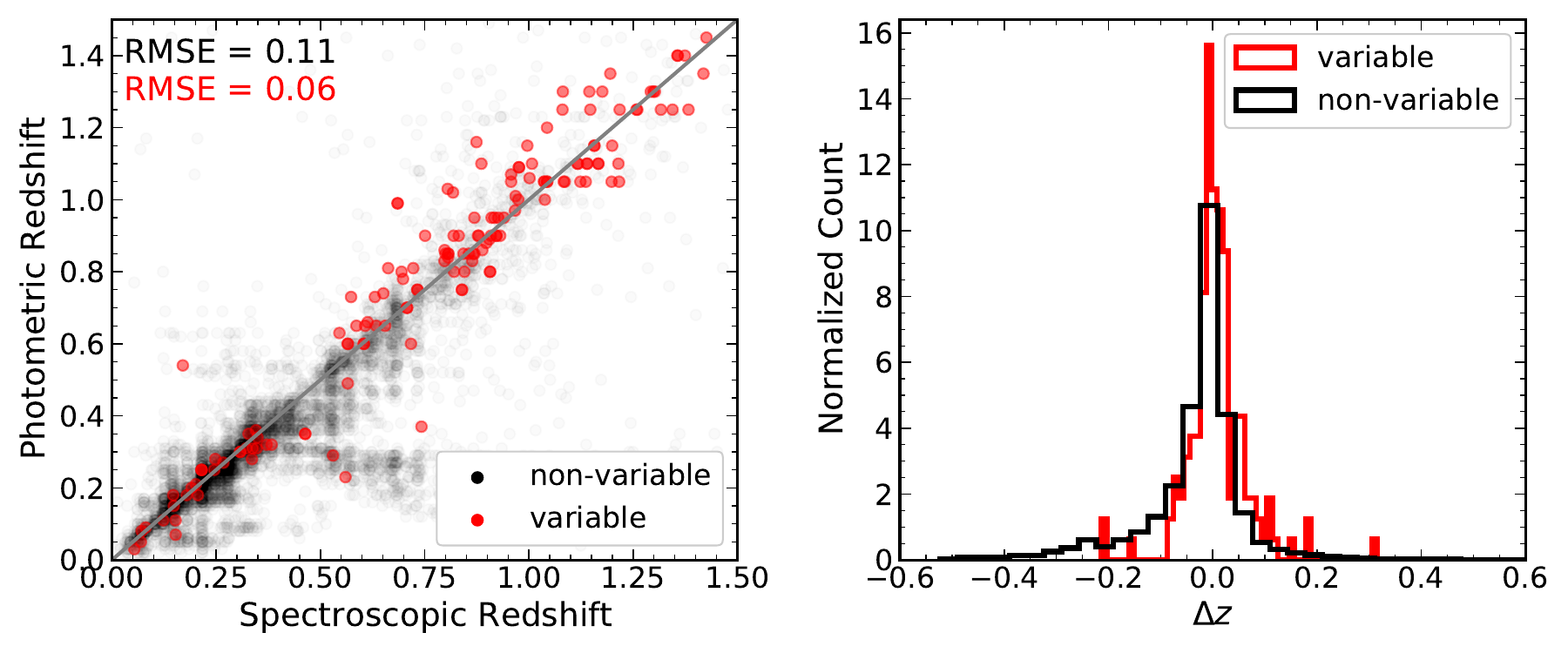}
\caption{Photometric redshifts versus matched sources with secure spectroscopic redshifts. The root mean square error (RMSE) for the variable (red) and non-variable (black) sources are shown in the upper-left hand corner of the figure. \label{fig:ozdes}}
\end{figure*}

Finally, we use the available spectroscopic redshifts of our deep field DES sources from \citet{Hartley2020} (see their Table~5) to benchmark our photometric redshifts (Figure~\ref{fig:ozdes}). For the photometric redshift comparison, we compute RMSE($\Delta z$), where $\Delta z = |z_{\rm{ph}} - z_{\rm{sp}}|/(1 + z_{\rm{sp}})$. Here, $z_{\rm{ph}}$ denotes our photometric redshift estimate and $z_{\rm{sp}}$ denotes the spectroscopic redshift. We find a RMSE of \RMSEznonvar{} for the non-variable galaxies and \RMSEzvar{} for variable galaxies. The details of the photometric redshift technique are given in \citet{Yang2017}.

\section*{Affiliations}
\noindent
\footnotesize
{\it
$^{1}$Department of Astronomy, University of Illinois at Urbana-Champaign, 1002 West Green Street, Urbana, IL 61801, USA\\
$^{2}$Center for AstroPhysical Surveys, National Center for Supercomputing Applications, 1205 West Clark Street, Urbana, IL 61801, USA\\
$^{3}$National Center for Supercomputing Applications, 1205 West Clark Street, Urbana, IL 61801, USA\\
$^{4}$Department of Astronomy, University of Geneva, ch. d’Ecogia 16, CH-1290 Versoix, Switzerland\\
$^{5}$Jodrell Bank Centre for Astrophysics, School of Physics \& Astronomy, The University of Manchester, Manchester M13 9PL, UK\\
$^{6}$Fermi National Accelerator Laboratory, P. O. Box 500,
Batavia, IL 60510, USA\\
$^{7}$Kavli Institute for Cosmological Physics, University of Chicago, Chicago, IL 60637, USA\\
$^{8}$Department of Physics and Astronomy, 4129 Frederick Reines Hall, University of California, Irvine, CA, 92697-4575, USA\\
$^{9}$Department of Physics, University of Illinois at Urbana-Champaign, 1110 West Green Street, Urbana, IL 61801, USA\\
$^{10}$Department of Astronomy \& Astrophysics, The
University of Chicago, Chicago IL 60637, USA\\
$^{11}$Department of Physics and Astronomy, Pevensey Building, University of Sussex, Brighton, BN1 9QH, UK\\
$^{12}$The Research School of Astronomy and Astrophysics,
Australian National University, ACT 2601, Australia\\
$^{13}$Center for Cosmology and Astro-Particle Physics, The Ohio State University, Columbus, OH 43210, USA\\
$^{14}$Kavli Institute for Particle Astrophysics \& Cosmology, P. O. Box 2450, Stanford University, Stanford, CA 94305, USA\\
$^{15}$Brookhaven National Laboratory, Bldg 510, Upton, NY 11973,
USA\\
$^{16}$ Laborat\'orio Interinstitucional de e-Astronomia - LIneA, Rua Gal. Jos\'e Cristino 77, Rio de Janeiro, RJ - 20921-400, Brazil\\
$^{17}$ Instituto de F\'{i}sica Te\'orica, Universidade Estadual Paulista, S\~ao Paulo, SP - 01140-070, Brazil\\
$^{18}$ Institute of Cosmology and Gravitation, University of Portsmouth, Portsmouth, PO1 3FX, UK\\
$^{19}$ CNRS, UMR 7095, Institut d'Astrophysique de Paris, F-75014, Paris, France\\
$^{20}$ Sorbonne Universit\'es, UPMC Univ Paris 06, UMR 7095, Institut d'Astrophysique de Paris, F-75014, Paris, France\\
$^{21}$ Department of Physics \& Astronomy, University College London, Gower Street, London, WC1E 6BT, UK\\
$^{22}$ Institut de F\'{\i}sica d'Altes Energies (IFAE), The Barcelona Institute of Science and Technology, Campus UAB, 08193 Bellaterra (Barcelona) Spain\\
$^{23}$ University of Nottingham, School of Physics and Astronomy, Nottingham NG7 2RD, UK\\
$^{24}$ Astronomy Unit, Department of Physics, University of Trieste, via Tiepolo 11, I-34131 Trieste, Italy\\
$^{25}$ INAF-Osservatorio Astronomico di Trieste, via G. B. Tiepolo 11, I-34143 Trieste, Italy\\
$^{26}$ Institute for Fundamental Physics of the Universe, Via Beirut 2, 34014 Trieste, Italy\\
$^{27}$ Observat\'orio Nacional, Rua Gal. Jos\'e Cristino 77, Rio de Janeiro, RJ - 20921-400, Brazil\\
$^{28}$ Department of Physics, University of Michigan, Ann Arbor, MI 48109, USA\\
$^{29}$ Hamburger Sternwarte, Universit\"{a}t Hamburg, Gojenbergsweg 112, 21029 Hamburg, Germany\\
$^{30}$ School of Mathematics and Physics, University of Queensland,  Brisbane, QLD 4072, Australia\\
$^{31}$ Centro de Investigaciones Energ\'eticas, Medioambientales y Tecnol\'ogicas (CIEMAT), Madrid, Spain\\
$^{32}$ Department of Physics, IIT Hyderabad, Kandi, Telangana 502285, India\\
$^{33}$ Santa Cruz Institute for Particle Physics, Santa Cruz, CA 95064, USA\\
$^{34}$ Institute of Theoretical Astrophysics, University of Oslo. P.O. Box 1029 Blindern, NO-0315 Oslo, Norway\\
$^{35}$ Instituto de Fisica Teorica UAM/CSIC, Universidad Autonoma de Madrid, 28049 Madrid, Spain\\
$^{36}$ Institut d'Estudis Espacials de Catalunya (IEEC), 08034 Barcelona, Spain\\
$^{37}$ Institute of Space Sciences (ICE, CSIC),  Campus UAB, Carrer de Can Magrans, s/n,  08193 Barcelona, Spain\\
$^{38}$ Faculty of Physics, Ludwig-Maximilians-Universit\"at, Scheinerstr. 1, 81679 Munich, Germany\\
$^{39}$ Center for Cosmology and Astro-Particle Physics, The Ohio State University, Columbus, OH 43210, USA\\
$^{40}$ Department of Physics, The Ohio State University, Columbus, OH 43210, USA\\
$^{41}$ Center for Astrophysics $\vert$ Harvard \& Smithsonian, 60 Garden Street, Cambridge, MA 02138, USA\\
$^{42}$ Australian Astronomical Optics, Macquarie University, North Ryde, NSW 2113, Australia\\
$^{43}$ Lowell Observatory, 1400 Mars Hill Rd, Flagstaff, AZ 86001, USA\\
$^{44}$ George P. and Cynthia Woods Mitchell Institute for Fundamental Physics and Astronomy, and Department of Physics and Astronomy, Texas A\&M University, College Station, TX 77843,  USA\\
$^{45}$ Instituci\'o Catalana de Recerca i Estudis Avan\c{c}ats, E-08010 Barcelona, Spain\\
$^{46}$ Physics Department, 2320 Chamberlin Hall, University of Wisconsin-Madison, 1150 University Avenue Madison, WI  53706-1390\\
$^{47}$ Institute of Astronomy, University of Cambridge, Madingley Road, Cambridge CB3 0HA, UK\\
$^{48}$ Department of Astrophysical Sciences, Princeton University, Peyton Hall, Princeton, NJ 08544, USA\\
$^{49}$ SLAC National Accelerator Laboratory, Menlo Park, CA 94025, USA\\
$^{50}$ School of Physics and Astronomy, University of Southampton,  Southampton, SO17 1BJ, UK\\
$^{51}$ Computer Science and Mathematics Division, Oak Ridge National Laboratory, Oak Ridge, TN 37831\\
$^{52}$ Department of Physics, Stanford University, 382 Via Pueblo Mall, Stanford, CA 94305, USA\\
$^{53}$ Kavli Institute for Particle Astrophysics \& Cosmology, P. O. Box 2450, Stanford University, Stanford, CA 94305, USA\\
$^{54}$ Max Planck Institute for Extraterrestrial Physics, Giessenbachstrasse, 85748 Garching, Germany\\
$^{55}$ Universit\"ats-Sternwarte, Fakult\"at f\"ur Physik, Ludwig-Maximilians Universit\"at M\"unchen, Scheinerstr. 1, 81679 M\"unchen, Germany\\
}

\bsp	
\label{lastpage}
\end{document}